\documentclass[twocolumn,times]{aastex62}
\usepackage{graphicx}
\usepackage[flushleft]{threeparttable}
\usepackage{blindtext}
\usepackage{ams math}
\usepackage{mathtools}
\usepackage{multirow}
\usepackage{hyperref}

\newcommand{\lya}{Ly$\alpha$ }

\begin{document}
\shorttitle{L\lowercase{y}${\alpha}$ EW distribution at $6.0 < z < 7.0$}
\shortauthors{Jung et al.}
\def\nar{New Astron.}
\def\na{New Astron.}
\title{\large \textbf{Texas Spectroscopic Search for \lya Emission at the End of Reionization \\I. Constraining the \lya Equivalent Width Distribution at 6.0 $<$ \textit{z} $<$ 7.0}}

\submitjournal{\textit{the Astrophysical Journal}}

\correspondingauthor{Intae Jung}
\email{itjung@astro.as.utexas.edu}

\author[0000-0003-1187-4240]{Intae Jung}
\affil{Department of Astronomy, The University of Texas at Austin, Austin, TX 78712, USA}

\author[0000-0001-8519-1130]{Steven L. Finkelstein}
\affil{Department of Astronomy, The University of Texas at Austin, Austin, TX 78712, USA}

\author[0000-0003-4456-1566]{Rachael C. Livermore}
\affil{Department of Astronomy, The University of Texas at Austin, Austin, TX 78712, USA}
\affil{School of Physics, University of Melbourne, VIC 3010, Australia}

\author[0000-0001-5414-5131]{Mark Dickinson}
\affil{National Optical Astronomy Observatory, Tucson, AZ 85719, USA}

\author[0000-0003-2366-8858]{Rebecca L. Larson}
\affil{Department of Astronomy, The University of Texas at Austin, Austin, TX 78712, USA}

\author[0000-0001-7503-8482]{Casey Papovich}
\affil{Department of Physics and Astronomy, Texas A\&M University, College
Station, TX, 77843-4242 USA}
\affil{George P.\ and Cynthia Woods Mitchell Institute for
  Fundamental Physics and Astronomy, Texas A\&M University, College
  Station, TX, 77843-4242 USA}

\author[0000-0002-8442-3128]{Mimi Song}
\affil{Astrophysics Science Division, Goddard Space Flight Center, Code 665, Greenbelt, MD 20771, USA}

\author[0000-0001-8514-7105]{Vithal Tilvi}
\affil{School of Earth \& Space Exploration, Arizona State University, Tempe, AZ 85287, USA}

\author[0000-0002-0784-1852]{Isak Wold}
\affil{Department of Astronomy, The University of Texas at Austin, Austin, TX 78712, USA}

%%%%%Abstract%%%%%
\begin{abstract}
The distribution of \lya emission is an presently accessible method for studying the state of the intergalactic medium (IGM) into the reionization era.  We carried out deep spectroscopic observations in order to search for \lya emission from galaxies with photometric redshifts $z=5.5-8.3$ selected from the Cosmic Assembly Near-infrared Deep Extragalactic Legacy Survey (CANDELS).  Utilizing data from the Keck/DEIMOS spectrograph, we explore a wavelength coverage of \lya emission at $z \sim 5-7$ with four nights of spectroscopic observations for 118 galaxies, detecting five emission lines with $\sim5\sigma$ significance: three in the GOODS-N and two in the GOODS-S field.  We constrain the equivalent width (EW) distribution of \lya emission by comparing the number of detected objects with the expected number constructed from detailed simulations of mock emission lines that account for the observational conditions (e.g., exposure time, wavelength coverage, and sky emission) and galaxy photometric redshift probability distribution functions.  The \lya EW distribution is well described by an exponential form, $\text{dN/dEW}\propto \text{exp(-EW/}W_0)$, characterized by the $e$-folding scale ($W_0$) of $\sim60-100$\AA\ at $0.3<z<6$.  By contrast, our measure of the \lya EW distribution at $6.0<z<7.0$ rejects a \lya EW distribution with $W_0 > 36.4$\AA$\ $(125.3\AA) at 1$\sigma$ (2$\sigma$) significance.  This provides additional evidence that the EW distribution of \lya declines at $z>6$, suggesting an increasing fraction of neutral hydrogen in the IGM at that epoch.
\end{abstract}

\keywords{early universe --- galaxies: distances and redshifts --- galaxies: evolution --- galaxies: formation --- galaxies: high-redshift --- galaxies: intergalactic medium}
%%%%%Section1: Introduction%%%%%
\section{Introduction}
Reionization was the last major phase transition of the intergalactic medium (IGM), and scrutinizing the detailed evolution of the IGM is a key frontier in observational cosmology.  High-redshift star-forming galaxies are thought to be the primary sources of ionizing photons \citep[e.g.,][]{Robertson2013a, Robertson2015a, Finkelstein2012a, Finkelstein2015a}.  Although bright quasi-stellar objects (QSOs) and active galactic nuclei (AGNs) are able to ionize their proximate areas as well and keep the IGM highly ionized during the last phase of cosmic reionization \citep[e.g.,][]{Giallongo2015a, Worseck2014a, DAloisio2017a, DAloisio2018a, Chardin2017a, Yoshiura2017a, Mitra2018a}, the number density of these objects rapidly decreases in the early universe.

Ionizing photon production and the escape fraction ($f_{\text{esc}}$) of these photons are key to modeling cosmic reionization.  The ionizing photon budgets are estimated from the cosmic star-formation history, which is tied to the shape of the galaxy luminosity function \citep[e.g.,][]{Bouwens2015a, Finkelstein2012a, Finkelstein2015a, Livermore2017a}, while the interstellar medium (ISM) and circumgalactic medium (CGM) determine $f_{\text{esc}}$ \citep[e.g.,][]{Paardekooper2015a, Kakiichi2017a, Katz2018a, Kimm2013a, Kimm2014a, Kimm2017a, Laursen2011a, Mason2017a, Rosdahl2018a}.  Thus, the interactions between high-redshift galaxies and the IGM have a significant impact on the evolution of the galaxies, and revealing detailed timelines of cosmic reionization and investigating IGM properties is key to gaining knowledge of galaxy evolution in the early Universe.

\textit{Wilkinson Microwave Anisotropy Probe (WMAP)} and \textit{Planck} observations constrain the midpoint of reionization to be $z\sim8-9$ \citep{Larson2011a, Planck-Collaboration2016a} from the measure of the large-scale polarization of the cosmic microwave background (CMB), while quasar observations studying the \lya forest and Gunn-Peterson effects at high redshift suggest a highly-ionized IGM at $z\sim6$ \citep[e.g.,][]{Becker2001a, Fan2006a, Bolton2011a, Mortlock2011a, McGreer2015a}.  Complementary measurements of the end of reionization based on the \lya emitter (LAE) luminosity function at $z\gtrsim6$ agree with those from the CMB and quasar observations \citep[e.g.,][]{Malhotra2004a, Ota2008a, Ota2017a, Ouchi2010a, Ouchi2018a, Zheng2017a}. However, robust studies of the IGM during reionization are still limited, as it is, for example, difficult to map the neutral fraction of the IGM during reionization with quasar spectroscopy due to the lack of a large population of quasars at $z>7$.

An immediately accessible method for studying the IGM in the reionization era is searching for \lya emission from continuum-selected galaxies with follow-up spectroscopy.  Due to the resonant nature of \lya scattering by neutral hydrogen, the presence of neutral hydrogen in the IGM easily attenuates Ly$\alpha$-emission-line strengths.  The fraction of continuum-selected Lyman break galaxies (LBGs) with strong spectroscopically-detected \lya emission (known as the ``\lya fraction'') was found to increase from $z =$ 3 to $z =$ 6 \citep{Stark2010a}.  It was thus expected that the \lya fraction at $z \sim$ 7 would be at least as high as at $z =$ 6 \citep{Stark2011a}.  However, initial studies have found an apparent deficit of strong \lya emission at $z >$ 6.5 \citep[e.g.,][]{Fontana2010a, Pentericci2011a, Pentericci2014a, Curtis-Lake2012a, Mallery2012a, Caruana2012a, Caruana2014a, Finkelstein2013a, Ono2012a, Schenker2012a, Schenker2014a, Treu2012a, Treu2013a, Tilvi2014a, Schmidt2016a, Vanzella2014a}.  The dust content of UV-selected galaxies has been found to decrease with increasing redshift \citep{Finkelstein2012b, Bouwens2014a, Marrone2018a}, thus the increased fraction of strong \lya emission from $z=3\rightarrow6$ is likely due to decreasing dust attenuation in galaxies.  A joint effect from metal poor stellar populations in the galaxies at higher redshift is likely as well, as it fosters the escape of \lya photons by enlarging HII regions from the generation of hard ionizing photons \citep[e.g.,][]{Finkelstein2011a, Nakajima2013a, Song2014a, Trainor2016a}.  Therefore, the perceived drop in \lya emission at $z>6$ is unlikely due to dust and implies that the neutral hydrogen fraction in the IGM increases significantly from $z\sim6$ $\rightarrow$ 7, although other galaxy evolutionary features and the uncertainties of the Lyman continuum escape fraction and the gas covering fraction need to be taken into account \citep[see][]{Papovich2011a, Finkelstein2012b, Dijkstra2014b}.

Despite this tantalizing evidence, measuring the \lya fraction depends on the sensitivity of the observed spectra and the completeness of the detected LAEs. \cite{De-Barros2017a} report a \lya fraction at $z\sim6$ lower than the values previously reported in the literature from a large sample of LAEs, and \cite{Caruana2018a} find no dependence of the \lya fraction on redshift at $3<z<6$ based on the analysis of 100 LAEs from the Multi-Unit Spectroscopic Explorer (MUSE)-Wide survey \citep{Herenz2017a}.

As discussed above, while many previous studies have used the \lya fraction as a measure of the evolution of \lya emission, it is a somewhat less constraining measure since it often does not account for the continuum luminosity of the host galaxy.  For this reason, we implement a more detailed analysis of our dataset, where we place constraints on the evolution of the \lya equivalent width (EW) distribution, using detailed simulations to include the effects of incompleteness.  This distribution function has been well studied at $0.3 < z < 6$, and has been found to have the form of an exponential distribution, $\text{dN/dEW}\propto \text{exp(-EW/}W_0)$, with a characteristic EW $e$-folding scale ($W_0$) of $\sim60\text{\AA}$ over the epoch $0.3 < z < 3$ \citep[e.g.,][]{Gronwall2007a, Guaita2010a, Ciardullo2012a, Wold2014a, Wold2017a}; and possibly higher at higher redshift \citep[e.g.,][]{Ouchi2008a, Zheng2014a}.  Particularly, in the epoch of reionization, the neutral hydrogen atoms in the IGM are expected to diminish these EWs, lowering the $e$-folding scale ($W_0$) of the observed \lya EW distribution \citep[e.g.,][]{Bolton2013a,Mason2017a}.  This gives us our research question: at what confidence can new observations rule out the $e$-folding scale ($W_0$) of $\sim60\text{\AA}$?  More importantly, understanding the evolution of the $e$-folding scale is key to predicting the number of \lya emitting galaxies with a given \lya EW distribution.

\begin{figure*}[t]
\centering
\includegraphics[width=0.57\paperwidth]{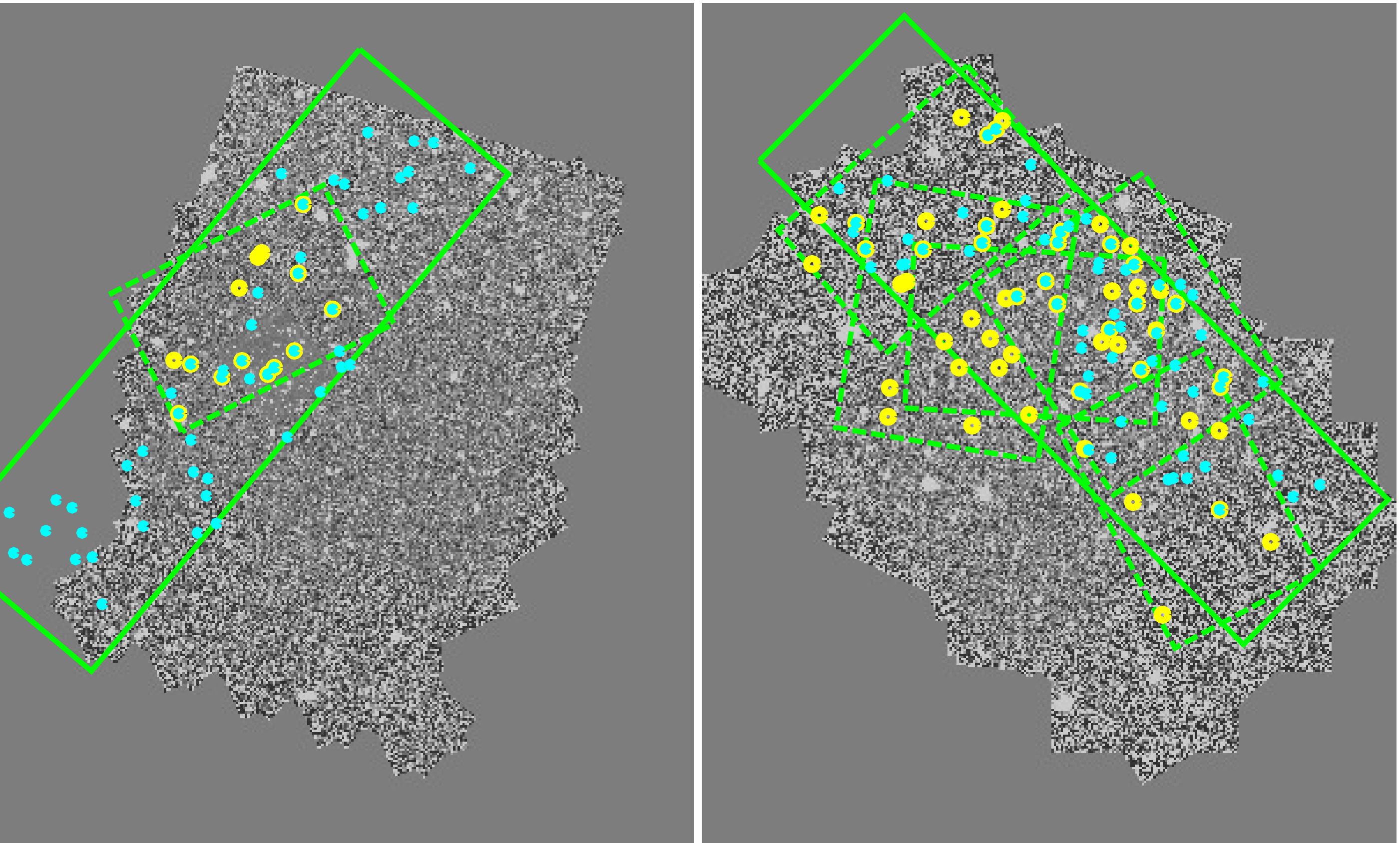}
\caption{Mask designs of our DEIMOS and MOSFIRE configurations overlaid in the GOODS-S (left) and GOODS-N (right) WFC3/F160W CANDELS images.  Observed areas are marked by green rectangles: larger solid rectangles (5$\arcmin$$\times$16.7$\arcmin$) show DEIMOS observations, and smaller dashed squares (6$\arcmin$$\times$4$\arcmin$) represent MOSFIRE observations.  Cyan and yellow circles are observed galaxies with DEIMOS and MOSFIRE, respectively.  While this figure shows our entire spectroscopic survey program with DEIMOS and MOSFIRE, we discuss our analysis with DEIMOS in this paper, and our follow-up paper will include the MOSFIRE data. }
\label{fig:mask_design}
\end{figure*}

To place observational constraints on the \lya EW $e$-folding scale, we performed Keck/DEIMOS (optical) spectroscopic observations of \lya emitting galaxies using a robust sample of candidate galaxies with photometric redshifts $z=5.5-8.3$.  We comprehensively accounted for incompleteness due to the noise level in the data (from a combination of telescope$+$instrument throughput, and also integration time) and the night sky lines which are ubiquitous at these wavelengths, and also due to galaxy photometric redshift probability distribution functions.

In this paper, we present our measure of the $e$-folding scale of the \lya EW distribution at $6.0<z<7.0$ measured from Keck/DEIMOS spectra. We describe our spectroscopic data in Section 2, and the detected emission lines are summarized in Section 3. Our analysis of the \lya EW distribution at $z\sim6.5$ is explained in Section 4.  Section 5 discusses the redshift dependence of the \lya EW $e$-folding scale, while our findings are summarized in Section 6.  We assume the \textit{Planck} cosmology \citep{Planck-Collaboration2016a} in this study, with $H_0$ = 67.8\,km\,s$^{-1}$\,Mpc$^{-1}$, $\Omega_{\text{M}}$ = 0.308 and $\Omega_{\Lambda}$ = 0.692, and a \cite{Salpeter1955a} initial mass function with lower and upper stellar-mass limits of 0.1 to 100 $M_{\odot}$ is assumed.  The \textit{Hubble Space Telescope (HST)} F435W, F606W, F775W, F814W, F850LP, F098M, F105W, F125W, F140W and F160W bands are referred as $B_{435}$, $V_{606}$, $i_{775}$, $I_{814}$, $z_{850}$, $Y_{098}$, $Y_{105}$, $J_{125}$, $JH_{140}$ and $H_{160}$, respectively.  All magnitudes are given in the AB system \citep{Oke1983a}.  All errors presented in this paper represent 1$\sigma$ uncertainties (or central 68\% confidence ranges), unless stated otherwise.

%%%%%Section2: Data%%%%%
\section{Data}
\subsection{Spectroscopic Survey and Sample Selection}
The target galaxies were selected from \cite{Finkelstein2015a} which found a sample of 7446 high-redshift candidate galaxies at $3.5<z<8.5$, using a photometric redshift ($z_{\text{phot}}$) measurement technique, in the CANDELS GOODS-South and -North fields.  We note that our entire observing program, the \textit{Texas Spectroscopic Search for \lya Emission at the End of Reionization}, utilizes 10 nights of MOSFIRE (near-infrared) observations as well as 4 nights of DEIMOS (optical) observations on the Keck telescopes.  Observations were conducted from Apr 2013 to Feb 2015, and two \lya detections at $z>7.5$ from the MOSFIRE observations are already published in \cite{Finkelstein2013a} and \cite{Song2016b}.  In our survey program, observations with both instruments (optical + near-infrared) enable us to put strong constraints on the observability of \lya emission, covering the broad range of galaxy photometric redshift probability distributions of $z\sim7$ candidate galaxies, as the DEIMOS and MOSFIRE combined wavelength coverage corresponds to redshifted \lya emission at $5<z<8$.  We first present the analysis of our DEIMOS observations in this paper, focusing on a \lya emission search at $5<z<7$, and a follow-up paper will include the MOSFIRE data to provide a comprehensive analysis, covering the entire wavelength of \lya emission at $z\sim5-8$.  

\begin{figure}[t]
\centering
\includegraphics[width=\columnwidth]{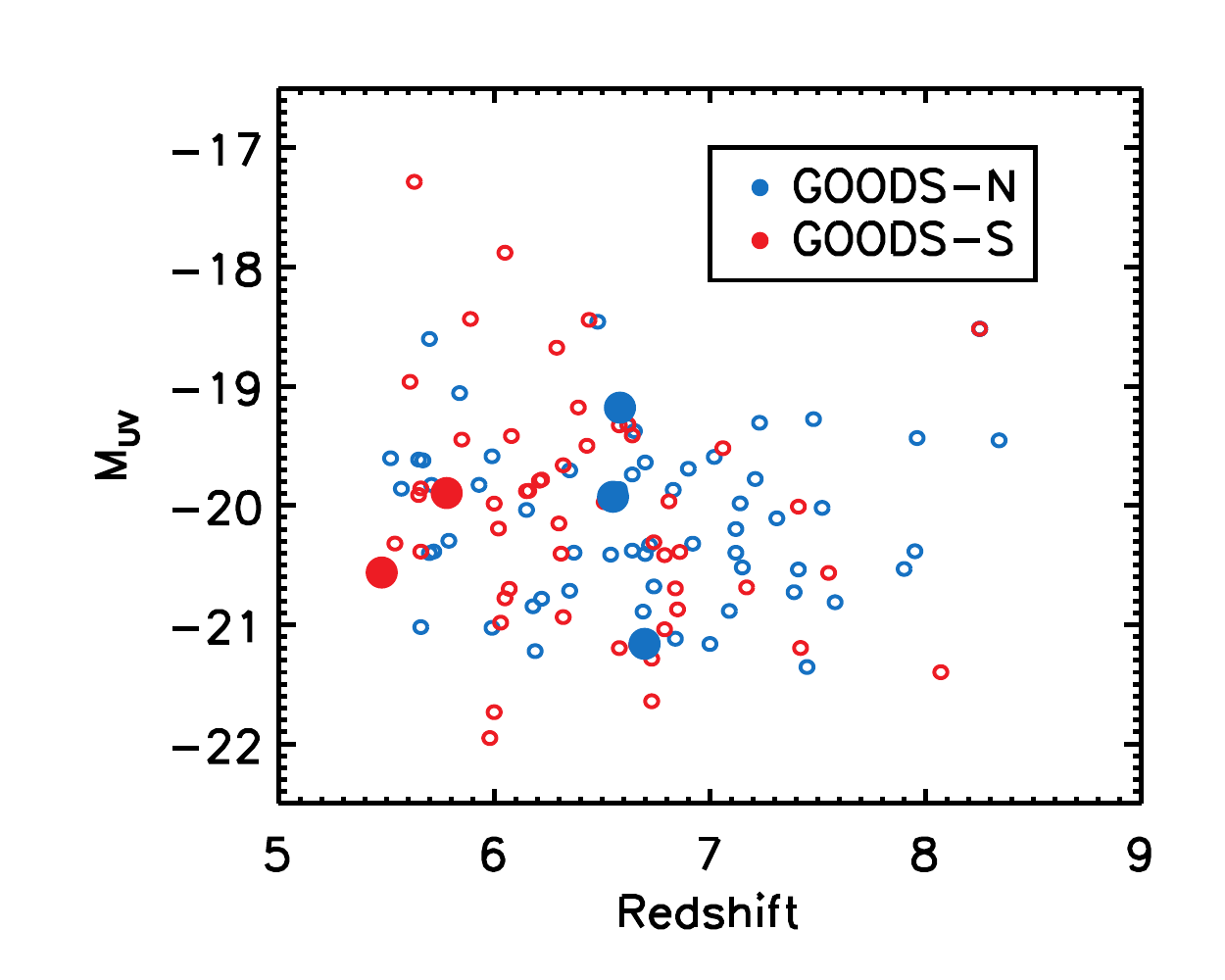}
\caption{$M_{\text{UV}}$ distribution of target galaxies in our DEIMOS observations as a function of redshift.  Open circles are target galaxies in GOODS-S (red) and GOODS-N (blue), and larger filled circles are \lya detections.  The line detections are discussed further in Section 3.}
\label{fig:muv_z}
\end{figure}

Figure \ref{fig:mask_design} shows the entire spectroscopic survey program.  The DEIMOS mask designs are solid rectangles, and the MOSFIRE masks are shown as dashed squares.  Cyan and yellow circles represent our target galaxies, observed by DEIMOS and MOSFIRE, respectively.  We also display the rest-frame ultraviolet (UV) magnitude ($M_{\text{UV}}$) distribution of our objects on slits as a function of redshift in Figure \ref{fig:muv_z}.  The rest-frame $M_{\text{UV}}$ is obtained from \cite{Finkelstein2015a}, who derive $M_{\text{UV}}$ through galaxy spectral energy distribution (SED) fitting.  $M_{\text{UV}}$ is measured based on an continuum flux of the best-fit model SED, which is averaged over a 100\AA-wide top-hat window centered at rest-frame 1500\AA.  As shown in the figure, our targeted galaxies cover range of $-22.0 \lesssim M_{UV} \lesssim -18.5$, except for a couple of faint objects with $M_{UV} \gtrsim -18$ found at $z\lesssim6$.

\subsection{DEIMOS Spectroscopy and Data Reduction}
We carried out four nights of observations for 118 galaxies with DEIMOS on the Keck 2 telescope: two nights in GOODS-S and two nights in GOODS-N (PI: Rachael Livermore).  However, we would note that the second night of observation for GOODS-N had relatively bad seeing and throughput ($\lesssim 20\%$ of that from the first night), thus we do not include this data in our analysis.  Also, unfortunately, for our GOODS-S data, we could not achieve properly flux-calibrated spectra due to the bad observing conditions.  We discuss the issue later in this section and in Appendix.

We used the same slitmask for the two nights on each field.  The slitmasks were designed using DSIMULATOR, a DEIMOS mask design tool.  We used the OG550 filter/830G grating centered at 9000\AA, which effectively covers a wavelength range, $\lambda\sim7000-10000$\AA\ (corresponding to \lya emission at $5\lesssim z\lesssim7$), and the spectral resolution ($\lambda/\Delta\lambda$) is $\sim3600$ with a 1.0$\arcsec$ slit width.  We use an A-B-B'-A (0.0$\arcsec$, +1.0$\arcsec$, -1.0$\arcsec$, 0.0$\arcsec$) dither pattern to reduce systematics (e.g., improving sky subtraction) in the final combined spectra, and clean cosmic-rays and detector defects.  We targeted 52 objects ($z_{\text{phot}}\gtrsim5.5$) in GOODS-S and 66 in GOODS-N: 58 at $5.5<z_{\text{phot}}<6.5$, 54 at $6.5<z_{\text{phot}}<7.5$, and 6 at $7.5<z_{\text{phot}}<8.3$.  The total exposure times are 22860 seconds (6.35hrs) for GOODS-S and 14400 seconds (4hrs) for GOODS-N.

The \texttt{spec2d} IDL pipeline developed for the DEEP2 Redshift Shift Survey \citep{Cooper2012a, Newman2013a} is publicly available for DEIMOS data reduction.  However, the public pipeline has technical issues with our dithered observational data.  Thus, we first obtained the sets of reduced individual science frames for individual slit objects from the pipeline, and then performed post-processing manually.  The extracted frames from the pipeline are already flat-fielded, rectified, and response corrected.  Every object spectrum spans two of the CCDs (blue and red sides) in the DEIMOS 4$\times$2 CCD configuration, and the pipeline reduces the spectrum independently for the blue and red sides.  Taking the individual rectified science frames, we cleaned cosmic-rays (CRs) using the IDL procedure \texttt{L.A.Cosmic} \citep{van-Dokkum2001a}.  Sky backgrounds were calculated by averaging two surrounding science frames, and we subtracted the sky backgrounds from all science frames before combining the individual frames.  Similar to \cite{Kriek2015a}, when combining CR-cleaned and sky-subtracted science images, we measured relative weights through different science frames, based on the maximum fluxes estimated from Gaussian fitting for continuum sources (e.g., alignment stars).  One-dimensional (1D) spectra of objects were extracted from the fully-reduced and combined two-dimensional (2D) spectra with $\sim$1.2 $\times$ the mean Gaussian FWHM along the spatial direction ($\sim 1.0 ''$). This follows an optimal extraction scheme \citep{Horne1986a}, which includes a spatial weight using a Gaussian profile in addition to an inverse-variance weight.  Since DEIMOS is not equipped with an atmospheric dispersion compensator (ADC), differential refraction is problematic in cases where observations span a large airmass, causing up to a few pixels of offset in y-axis (spatial direction) on tracing object positions in the blue and red sides of 2D spectra \citep{Szokoly2005a, Newman2013a}.  To correct this in our observations, we calculate the spatial offset of our four guide stars between the expected and the actual locations, independently in the blue and red sides of 2D spectra.  Centering guide stars is done by Gaussian fitting along the spatial (y-axis) direction at every pixel in the wavelength direction (x-axis), and we measured the median offsets in the blue and red sides of the 2D spectra for all four guide stars. Lastly, by averaging the pre-obtained median offsets from the guide stars, we obtained the mean blue and red side spatial offsets, -0.66 and +0.64 pixels, respectively.  Thus, we applied these offsets on locating object spatial positions when we extract 1D spectra.

Flux calibration and telluric absorption correction was done by using the model stellar spectrum \citep{Kurucz1993a} that has the same spectral type (B2IV) to the standard star (BD+33d2642 for GOODS-N).  The standard star was observed in three science frames with a long slit and a 45-second exposure in each frame.  Data reduction was done in the same manner as was used for reducing the spectra of our science objects as described earlier.  The response profiles of the stellar 1D spectra were derived for both the blue and red sides separately, dividing the reduced 1D spectra of the standard star by the Kurucz model spectra.  Absolute flux calibration was done simultaneously by measuring a scaling of the spectra to match their known $z$-band magnitudes.  To test our flux calibration, we calculate $HST$/ACS $z_{850}$-band fluxes of mask alignment stars, and compare the measured fluxes to the $HST$ CANDELS imaging data.  Slit losses were taken into account in this procedure, assuming our target galaxies are point sources, as the high-redshift galaxies are unresolved under the seeing of our observations.  The estimated accuracy level of the flux calibration was $\sim20\%$ in flux (see Appendix), thus we include 20\% systematic errors in our emission line flux measurement.

As mentioned earlier, our flux calibration of GOODS-S data was less successful.  This is due to the bad observing conditions (a large variation of the seeing between science objects and standard star observations, and a large drift on centering 2D spectra due to atmospheric dispersion, which is problematic to center 2D spectra precisely in the spatial direction).  As the $z_{850}$ magnitudes of the GOODS-S continuum sources, which are calculated with the $HST$/ACS 850LP filter curve, are significantly different from those from \cite{Finkelstein2015a}, the calibrated fluxes for the GOODS-S dataset are unreliable.  Thus, we were unable to use the calibrated \lya line flux of the GOODS-S objects in our analysis in later sections. The details on flux calibration are discussed in Appendix.

\begin{table*}[t]
\centering
\begin{center}
\caption{Summary of Ly$\alpha$ Emitters Observed with Keck/DEIMOS}
\label{tab:candidates}
\begin{tabular}{ccccccccc}
\tableline 
\tableline
\quad {ID\tablenotemark{a}} & {R.A.} & {Decl.} & {$z_{\text{spec}}$} & {$z_{\text{phot}}$} & {$J_{125}$} & {$M_{\text{UV}}$} & {$f_{\text{\lya}}$} & {EW$_{\text{Ly}\alpha}$}  \\
\quad {}     & {(J2000.0)} & {(J2000.0)} & {}             & {}                           & {}                 &  {}                          & {($10^{-18}$erg s$^{-1}$cm$^{-2}$)}&  {(\AA)} \\
\tableline
\quad{z6\_GND\_28438} & {189.177979} & {62.223713} & {6.551 $\pm$ 0.002} & {6.12 $^{+0.21}_{-4.48}$} & {26.51 $^{+0.09}_{-0.08}$} & {-19.93} & {3.08 $\pm$ 0.56} & {20.50 $\pm$ 3.90}\\
\quad{z6\_GND\_5752} & {189.199585} & {62.320965} & {6.583 $\pm$ 0.004} & {5.70 $^{+0.32}_{-4.72}$} & {27.32 $^{+0.20}_{-0.17}$} & {-19.18} & {3.34 $\pm$ 0.67} & {36.72 $\pm$ 7.35}\\
\quad{z7\_GND\_10402} & {189.179276} & {62.275894} & {6.697 $\pm$ 0.001} & {7.00 $^{+0.17}_{-0.25}$} & {25.75 $^{+0.06}_{-0.06}$} & {-21.16} & {3.08 $\pm$ 0.59} & {16.86 $\pm$ 2.51}\\
\quad{z5\_MAIN\_4396\tablenotemark{b}} & {53.138580} & {-27.790218} & {5.479 $\pm$ 0.001} & {5.18 $^{+0.06}_{-0.08}$} & {25.93 $^{+0.02}_{-0.02}$} & {-20.56} & {-} & {-}\\
\quad{z6\_GSD\_10956} & {53.124886} & {-27.784111} & {5.780 $\pm$ 0.002} & {5.51 $^{+0.26}_{-0.21}$} & {26.78 $^{+0.12}_{-0.10}$} & {-19.90} & {-} & {-}\\
\tableline
\end{tabular}
\end{center}
\begin{flushleft}
\tablenotetext{a}{The object IDs are from \cite{Finkelstein2015a}, encoded with their photometric redshifts and the fields in the CANDELS imaging data.}
\tablenotetext{b}{\lya detection with $z=5.42\pm0.07$ for this object is reported in \cite{Rhoads2005a} from their $HST$/ACS grism survey.}
\end{flushleft}
\label{tab:LAEs}
\end{table*}

\begin{figure*}[t]
\centering
\includegraphics[width=0.84\paperwidth]{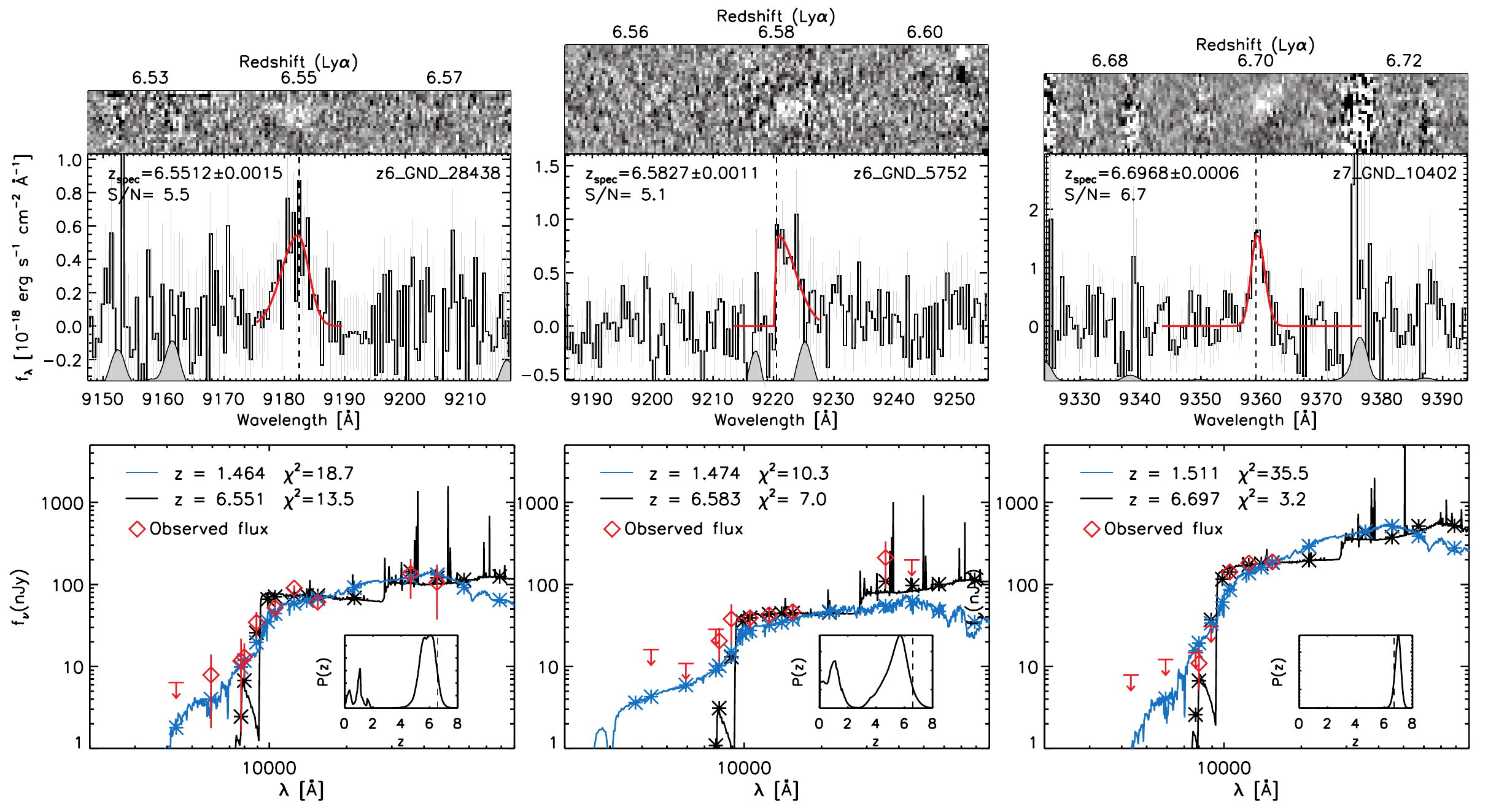}
\caption{(Top panels) One- and two-dimensional spectra of line-detected objects in GOODS-N. Red curves show the best-fit asymmetric Gaussian curves. (Bottom) Galaxy SED fitting results. Each panel shows two SEDs for high-$z$ (Ly$\alpha$) and low-$z$ (OII) solutions (black and blue solid curves, respectively). Red diamonds are observed fluxes with their associated errors. Photometric redshift probability distributions, $P(z)$ taken from \cite{Finkelstein2015a}, are displayed as inset figures, and the best-fit photometric redshifts are shown with vertical dashed lines.}
\label{fig:emissions1}
\end{figure*}

\begin{figure}[t]
\centering
\includegraphics[width=\columnwidth]{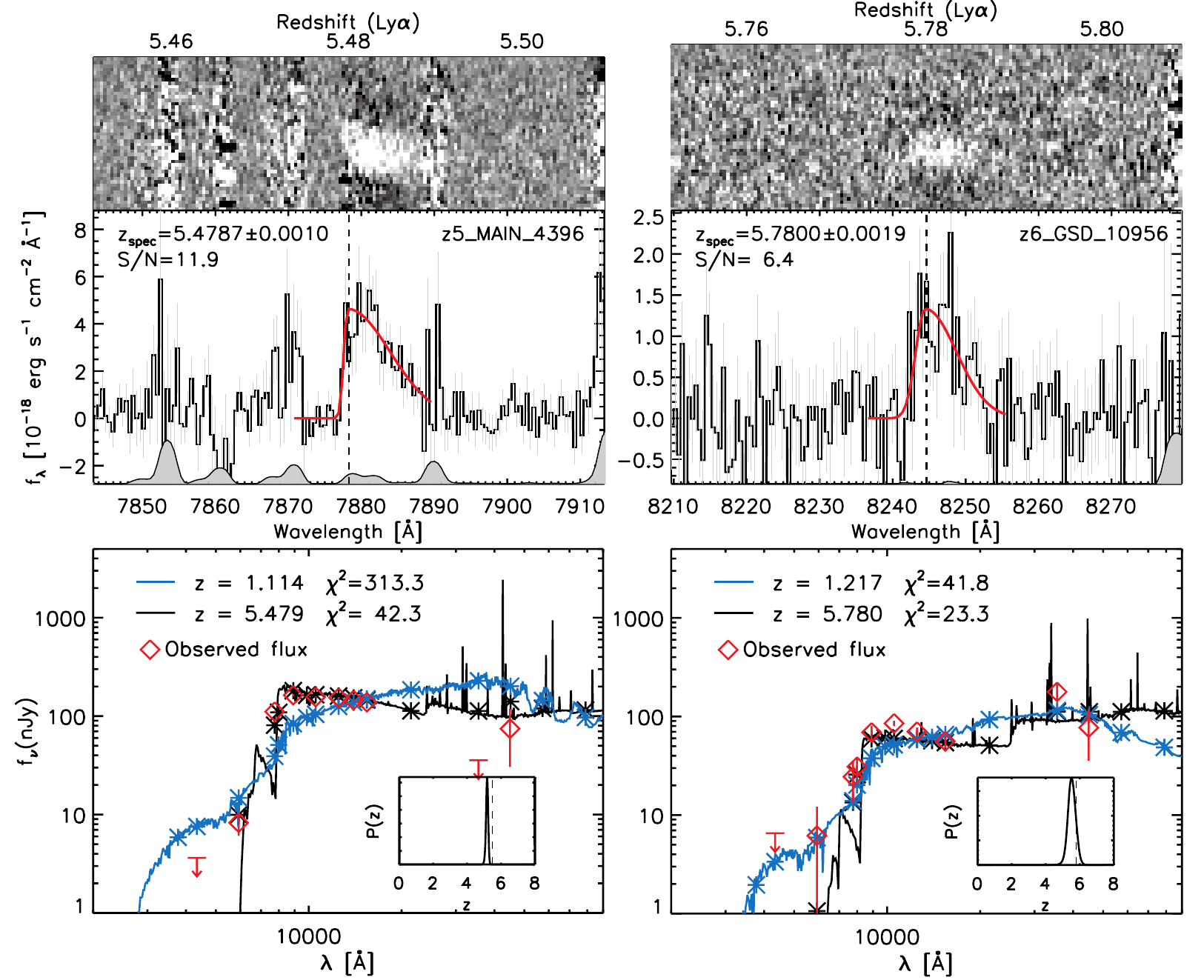}
\caption{Same as Figure \ref{fig:emissions1} but for the GOODS-S objects.}
\label{fig:emissions2}
\end{figure}

%%%%%Section3: Results%%%%%
\section{Results}
\subsection{Emission line detection}
We searched for emission lines utilizing both the 1D and 2D spectra.  After a primary visual inspection of the 2D spectra to search for significant lines, we generated a list of potential emission lines from an automated emission line search routine.  In this search we fit an asymmetric Gaussian function to 1D spectra at intervals of $\sim3$\AA, which is comparable to the instrumental spectral resolution, using \texttt{MPFIT} \citep{Markwardt2009a}.  We derive the signal-to-noise (S/N) levels of the machine-detected lines via 1000 Monte-Carlo (MC) simulations, modulating the 1D spectrum within the 1D noise level.  If any emission feature is detected with S/N $\gtrsim 4$, it is recorded in the list of potential emission lines.  Candidates among the machine-detected lines are confirmed via secondary visual inspection of the 2D spectra.  

With the systematic emission line search we find five (probable) \lya emission lines at $z\gtrsim5.5$.  The five LAEs are summarized in Table \ref{tab:LAEs}: three detections in GOODS-N and two in GOODS-S.  The object IDs in the table are encoded with their photometric redshifts and the fields where they are detected in imaging data. The 1D and 2D spectra of the LAEs are shown in Figure \ref{fig:emissions1} (GOODS-N) and Figure \ref{fig:emissions2} (GOODS-S).  We check if any of our detected LAEs are previously reported in the literature, and find that z5\_MAIN\_4396 in the Hubble Ultra Deep Field (HUDF) was previously detected as a \lya emitting galaxy with a measured \lya redshift \citep[$z=5.42\pm0.07$;][]{Rhoads2005a} from the $HST$/ACS grism survey program, the Grism ACS Program for Extragalactic Science (GRAPES) \citep[PI: S. Malhotra;][]{Pirzkal2004a}. Our spectroscopic redshift ($z=5.479$) is within their 1$\sigma$ error.  However, as the GOODS-S data cannot be properly flux-calibrated, the two emission galaxies (z5\_MAIN\_4396 and z6\_GSD\_10956) in GOODS-S are excluded from further analysis in this paper, which requires the calibrated \lya line flux. The line fluxes are measured from asymmetric Gaussian fitting (red curves in the figures).  For calculating the EW of the detected \lya line we use a continuum flux of the best-fit model galaxy SED, which is averaged over a 100\AA\ window redward of the \lya line. The model SEDs are constructed from the \cite{Bruzual2003a} stellar population synthesis model, and our SED fitting is further described in Section 3.2.

We check the possibility of the detected emission lines being low-$z$ contaminants.  First, to rule out the possibility of being [OII] $\lambda\lambda$3726, 3729, we compare the high-$z$ solutions of the \lya lines to galaxy SED fitting results at the redshift of [OII] (bottom panels in Figure \ref{fig:emissions1} and Figure \ref{fig:emissions2}).  In the case of z7\_GND\_10402 and z5\_MAIN\_4396, due to their strong Lyman-break our SED fitting strongly rejects the low-$z$ solutions, while for z6\_GND\_28438, z6\_GND\_5752, and z6\_GSD\_10956, we cannot rule out the low-$z$ solutions in the SED fitting results.  However, if the emission lines we find are one of the OII doublets, the DEIMOS spectral resolution, $\sim3$\AA, can distinguish the two peaks with a gap of $\sim$7--8\AA\ at $z\sim1.5$.  Inspecting the 2D spectral images of the two objects, we cannot find any significant second peak of the doublet nearby the detected emission line.  Although there are small bumps found in their 1D spectra at 9172\AA\ for z6\_GND\_28438 and at 9230\AA\ for z6\_GND\_5752 (Figure \ref{fig:emissions1}), those are not considerably favored as the line intensity ratios of the bumps to the detected emission lines (5.36 for z6\_GND\_28438 and 0.22 for z6\_GND\_5752) do not satisfy the physically motivated low- and upper-limits of the line intensity ratio of the OII doublet $I(3729)/I(3726)$, which is from 0.35 to 1.5 \citep{Pradhan2006a}.

We also visually inspect the 1D and 2D spectra of the emission galaxies to find any features of H$\beta$ and [OIII] $\lambda\lambda$4959, 5007.  In the case that any of our detected lines are one of H$\beta$ and the OIII doublet, the expected gaps between the lines range from 230 -- 282\AA\ at their low-$z$ solutions of $z=0.57$ -- 0.93.  We check the expected locations of the three emission lines (H$\beta$ and [OIII] $\lambda\lambda$4959, 5007) individually, assuming that the detected emission is one of the three lines.  We also check the possibility of being H$\alpha$ emission, searching for nearby emission features, [NII] $\lambda\lambda$6548, 6584.  From visual inspection, we find no significant emission features, although there were some cases that the expected line locations fall close to sky emission lines so that we are unable to completely ignore the possibility.  H$\beta$ and [OIII] $\lambda$4959 may be weaker than [OIII] $\lambda$5007, and [NII] $\lambda\lambda$6548, 6584 are weaker than H$\alpha$ as well, thus the expected S/N levels of the lines could be low, making it difficult to detect them.  Thus, we additionally compare galaxy SED fitting results with the low-$z$ solutions of [OIII] $\lambda$5007 and H$\alpha$ to those with the high-$z$ \lya solutions in order to check the low-$z$ possibility, but SED analysis alone cannot completely rule out the possibility.  Although we still cannot completely rule out all the scenarios of low-$z$ interlopers for our sample (except for z5\_MAIN\_4396) in the given S/N levels, the low-$z$ solutions are highly unlikely as discussed above (e.g., no significant spectral features found).  Therefore, we consider all the five detections as \lya in our further analysis.
\begin{table}[t]
\centering
\begin{center}
\caption{Summary of the Physical Properties of the Ly$\alpha$ Emitters}
\label{tab:SED_fitting}
\begin{tabular}{cccc}
\tableline 
\tableline
\quad {ID} & {log $M_{*}/M_{\odot}$} & {SFR ($M_{*}/M_{\odot}$)} & {$E(B-V)$}	 \\
\tableline
\quad {z6\_GND\_28438} & {9.43$^{+0.12}_{-0.17}$} & {22.1 $^{+9.5}_{-7.3}$} & {0.19 $^{+0.03}_{-0.04}$} \\
\quad {z6\_GND\_5752} & {9.30$^{+0.28}_{-0.29}$} & {11.6$^{+11.2}_{-5.6}$} & {0.16$^{+0.07}_{-0.06}$}\\
\quad {z7\_GND\_10402} & {9.42$^{+0.47}_{-0.43}$} & {16.7$^{+24.5}_{-10.2}$} & {0.31$^{+0.10}_{-0.10}$}\\
\quad {z5\_MAIN\_4396} & {8.78$^{+0.02}_{-0.01}$} & {12.4$^{+0.2}_{-0.2}$} & {0.004$^{+0.002}_{-0.002}$}\\
\quad {z6\_GSD\_10956} & {9.23$^{+0.09}_{-0.09}$} & {8.7$^{+1.9}_{-1.4}$} & {0.06$^{+0.02}_{-0.02}$}\\
\tableline
\end{tabular}
\end{center}
\tablecomments{Physical quantities of each object are derived from galaxy spectral energy distribution (SED) fitting with photometric data.}
\end{table}

\subsection{Galaxy physical properties}
To derive physical galaxy properties, we perform galaxy SED fitting with the $HST$/ACS (F435W, F606W, F775W, F814W and F850LP) + WFC3 (F105W, F125W, F140W and F160W) and $Spitzer$/IRAC 3.6$\mu$m and 4.5$\mu$m band fluxes of the line-detected galaxies.  We use the \lya emission line-subtracted fluxes in SED fitting, subtracting the measured \lya emission fluxes from the $z_{850}$ and $Y_{105}$ band continuum fluxes.  The details of our SED fitting are described in \cite{Jung2017a}. Briefly, it uses a Markov Chain Monte Carlo (MCMC) algorithm to fit the observed multi-wavelength photometric data with the model galaxy SEDs based on the \cite{Bruzual2003a} stellar population synthesis models.  The physical properties of the five LAEs derived from our SED fitting are summarized in Table \ref{tab:SED_fitting}, and the derived stellar masses and UV-corrected SFRs show that our LAEs are typical star-forming galaxies, distributed within the $\sim1\sigma$ scatter of the typical $M_{*}-$SFR relation at $z\sim6$ \citep{Salmon2015a, Jung2017a}.

%%%%%Section4: Measuring a LyA EW distribution%%%%%
\begin{figure}[t]
\centering
\includegraphics[width=\columnwidth]{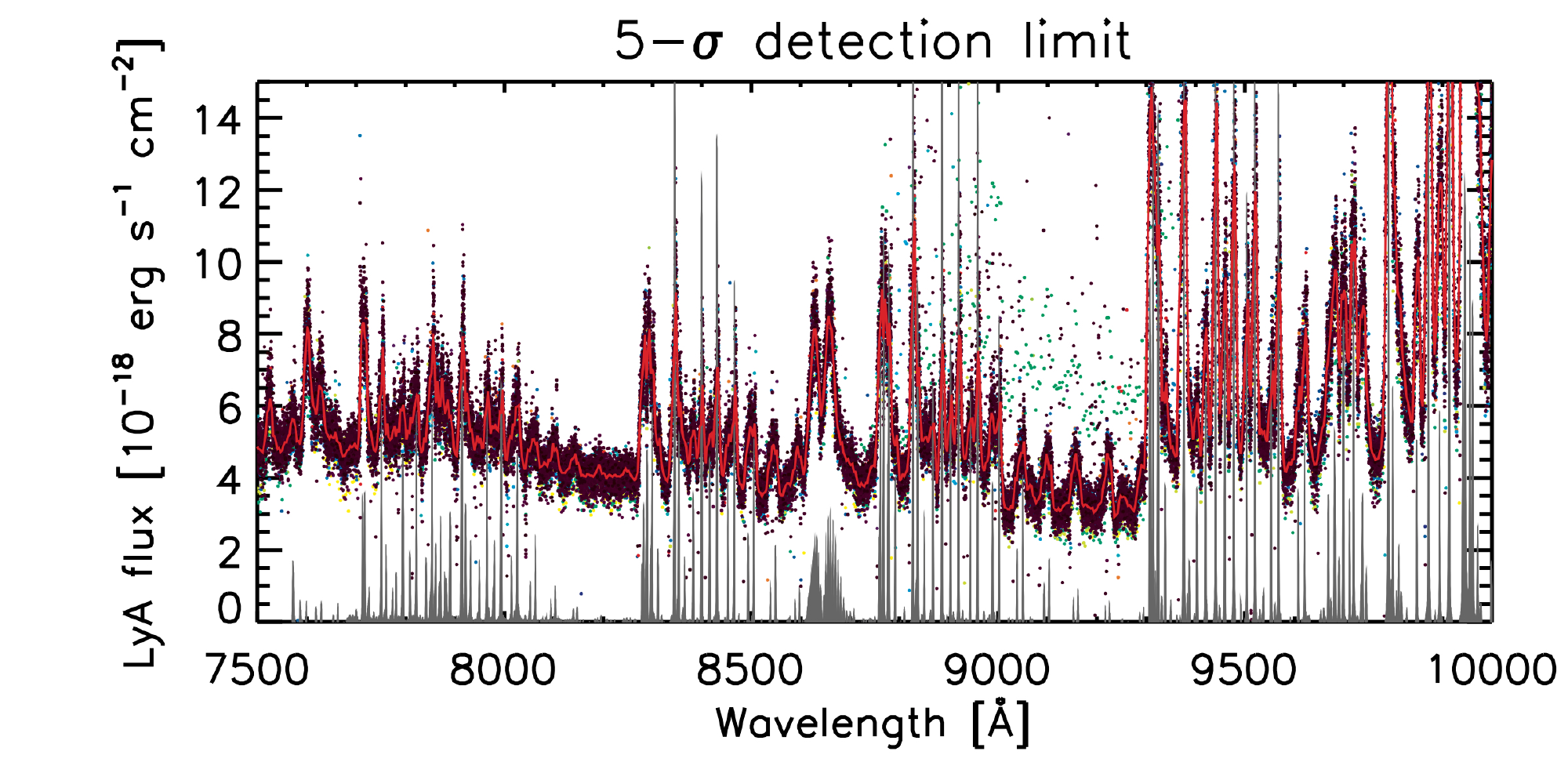}
\caption{The $5\sigma$ detection limit of an emission line flux across the instrument wavelength coverage, measured with 3\AA\ spacing using a Monte-Carlo simulation, inserting mock emission lines.  The colored dots show the measured detection limit from the different galaxies, and the median detection limit is drawn as the red curve.  On the bottom, grey shaded regions represent sky emission lines. Between the sky emission lines, the typical $5\sigma$ detection limit is $\sim$ 3--5 $\times 10^{-18}$ erg s$^{-1}$ cm $^{-2}$.  We derive a linear relation between the line strength and its S/N level across the instrument wavelength coverage, and the detection limit of each simulated \lya is interpolated from the pre-calculated linear relation.}
\label{fig:sensitivity}
\end{figure}

\section{Measuring the L\lowercase{y}$\alpha$ EW distribution}
The EW distribution of \lya emission is often described by an exponential form, $\text{dN/dEW}\propto \text{exp(-EW/}W_0)$, characterized by the $e$-folding scale, $W_0$ \cite[e.g.,][]{Cowie2010a}.  The measured $e$-folding scale at $0.3 < z < 3.0$ is $\sim 60$\AA\ \citep[e.g.,][]{Gronwall2007a, Nilsson2009a, Guaita2010a,  Blanc2011a, Ciardullo2012a, Wold2014a, Wold2017a}, and increases to higher redshift \citep[e.g.,][]{Zheng2014a, Hashimoto2017a}.  One would expect to see a relatively reduced EW of observed \lya emission from galaxies before cosmic reionization is completed, compared to the quantity observed from galaxies at lower redshift in the absence of galaxy evolution \citep{Bolton2013a, Mason2017a}, because \lya photons emitted from high-redshift galaxies are resonantly scattered by neutral hydrogen atoms in the IGM.  Thus, a measure of the $e$-folding scale of the \lya EW distribution at the end of reionization is a key observable, which reflects the ionization status of the IGM.  In this section, we provide our measure of this $e$-folding scale at $6.0<z<7.0$.

\begin{figure*}[t]
\includegraphics[width=1.05\columnwidth]{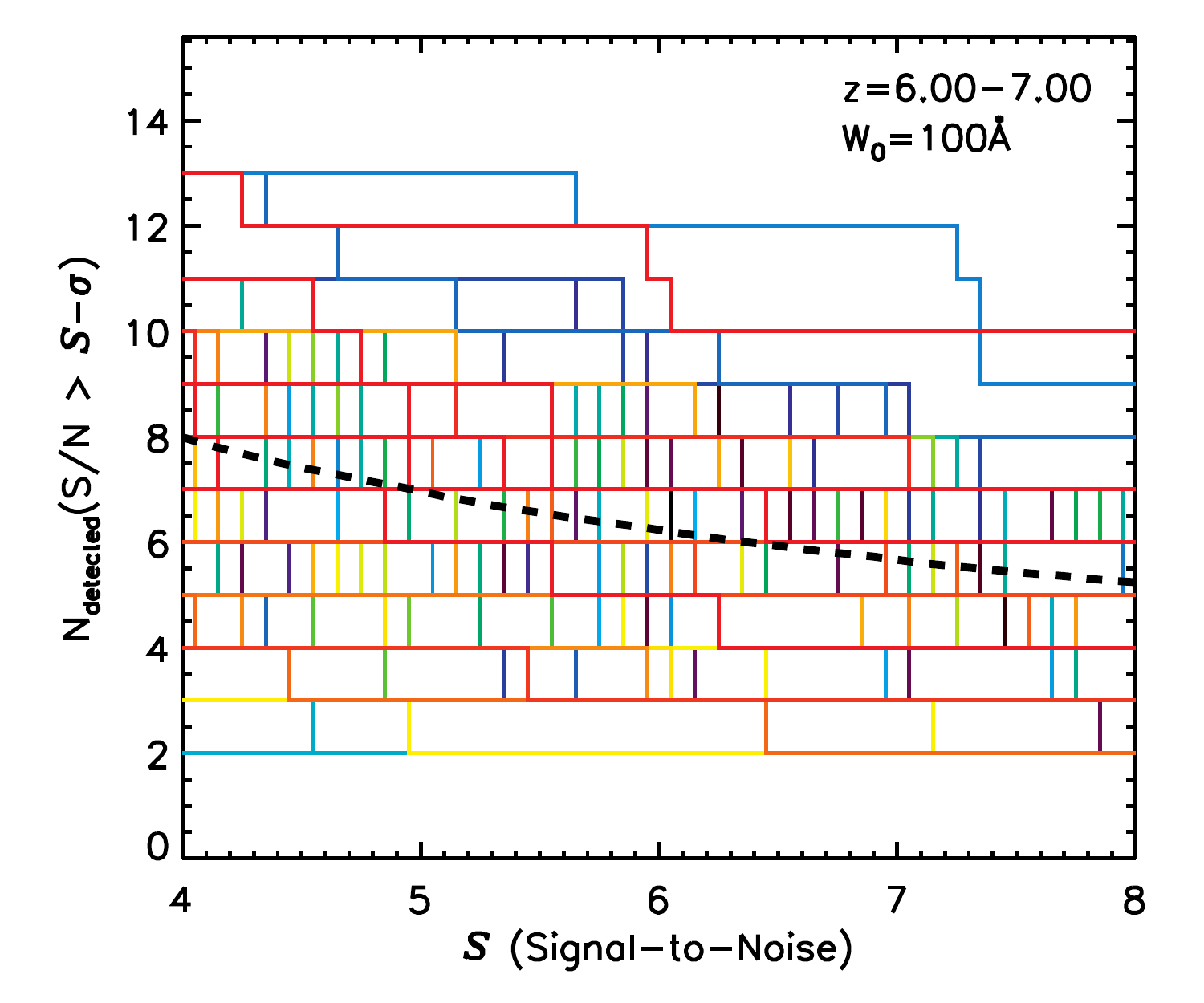}
\includegraphics[width=1.05\columnwidth]{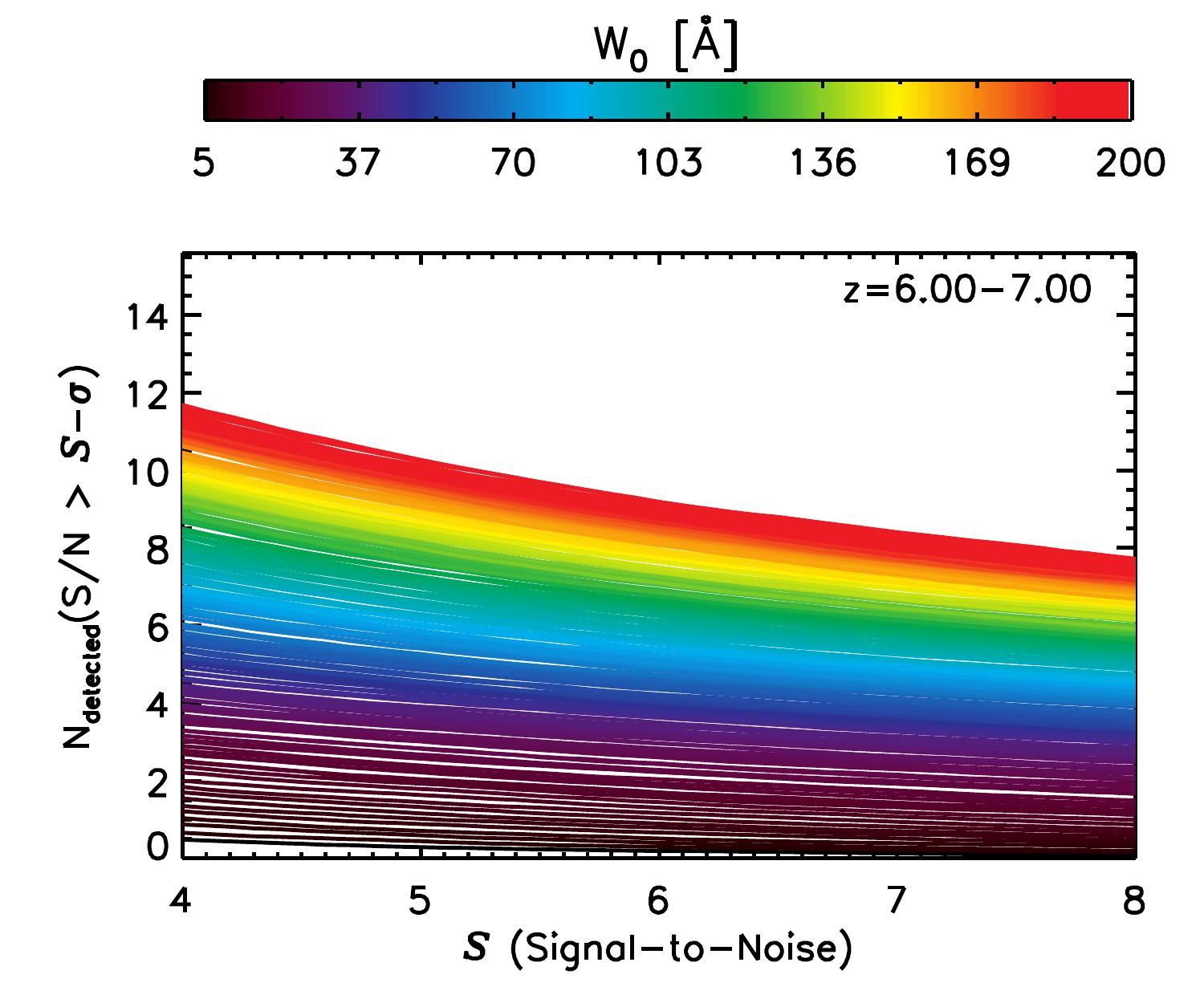}
\caption{(Left) 100 Monte-Carlo simulations of the expected number of detections as a function of S/N level ($\mathcal{S}$) with $W_0=100$\AA\ at $z\sim6.5$. We show only 100 of the 1000 simulation results for clarity.  Each simulation is denoted by a different color, and the dashed curve shows the mean value, averaged over the 1000 simulations. (Right) the expected number of detections as a function of S/N level ($\mathcal{S}$) with various EW distributions at $z\sim6.5$. A larger choice of the $e$-folding scale ($W_0$) of the \lya EW distribution (redder color) predicts a larger number of \lya detections.}
\label{fig:n_detection}
\end{figure*}

\subsection{Simulating the expected number of detections}
A simple but key observable from our spectroscopic survey is the number of detected \lya emission lines.  This depends not only on the observed \lya EW distribution of the observed galaxies, but also the completeness of the observations.  
In this study, we wish to test the hypothesis that a uniform quenching of \lya emission from a partially-neutral IGM is evolving the $e$-folding scale of the EW distribution towards lower values at $z\gtrsim6$.  To facilitate this, we develop simulations which assess the likelihood of detecting a \lya emission line of a given strength accounting for all sources of incompleteness (e.g., spectroscopic depth, sky lines, wavelength coverages, $P(z)$ distribution).  \cite{Song2016b} described this as a \lya visibility test, comparing the number of Ly$\alpha$ detections above a specific S/N level to that expected, with the latter calculated by assigning mock \lya emission profiles in 1D spectra of target galaxies in a Monte-Carlo fashion.  The \lya wavelengths were drawn from the photometric redshift probability distribution function, $P(z)$, and \cite{Song2016b} adopted the intrinsic EW distribution from \cite{Schenker2014a}, which is based on published data at $3<z<6$, when the IGM was ionized.
We advance the \lya visibility test of \cite{Song2016b} by setting the EW $e$-folding scale $W_0$ as a free parameter which we constrain with our observations. 

To estimate the expected S/N levels for the simulated \lya lines, we derive the detection limit for every individual galaxy at each wavelength by adding a mock emission line to the galaxy 1D spectra.  We assume this mock line has an intrinsic line profile equal to the best-fit asymmetric Gaussian profile of our highest S/N \lya emission detected in z5\_MAIN\_4396 (FWHM$_{\text{blue}}=0.88$\AA\ and FWHM$_{\text{red}}=9.69$\AA), obtained via \texttt{MPFIT}.  We add in this emission line at each wavelength step at a range of emission line fluxes, and then measure the resultant S/N level of each line in the same Monte Carlo fashion as done on the real lines.  We derive a linear relation between the line strength and its S/N level at all wavelengths with 3\AA$\ $spacing, a comparable size to the spectral resolution in our observational setting.  This measurement allows us to determine the expected S/N levels for a given \lya emission strength across the entire wavelength range.  A typical 5$\sigma$ detection limit of \lya flux is $\sim$3--5$\times$10$^{-18}$ erg s$^{-1}$ cm$^{-2}$ between sky emission lines as shown in Figure \ref{fig:sensitivity}.  To check any dependence of the derived \lya detection limit on the shape of the mock emission profile (specifically, FWHM), we re-did the simulations assuming a narrower mock line profile with $\text{FWHM}=5$\AA.  In \cite{Mallery2012a}, the size of the \lya FWHM at $3.8<z<6.5$ ranges from 5.71\AA\ to 10.88\AA\ (68\% confidence), so our tested range of FWHM from 5\AA\ to 10\AA\ reasonably considers the typical \lya line profile.  The smaller choice of $\text{FWHM}=5$\AA\ makes the line profile sharper, and slightly lowers the derived \lya line detection limit.  However, the overall difference of the estimated detection limit of an emission line flux between $\text{FWHM}=$5\AA\ and 10\AA\ is  below the $\sim10\%$ level. 

Using these S/N values, we then calculate the expected number of detections for a range of potential $W_0$ values.   In this simulation a Monte-Carlo aspect is needed, as the broad photometric redshift distributions require us to sample a broad wavelength range, and the line strength of the simulated \lya emission lines are sampled through the assumed EW distribution.  For each mock emission line, we i) assign a wavelength for the \lya line by drawing randomly from the photometric redshift distribution: $\lambda_{\text{\lya}}=(1+z)\times1215.67$\AA, ii) assign the line strength by drawing from the assumed \lya EW distribution: $P(\text{EW}) \propto \text{exp}^{-\text{EW}/W_0}$, which is based on the inferred continuum magnitudes near the wavelength of \lya (averaged over a 100\AA\ window redward of \lya emission), and iii) determine the S/N level of the simulated \lya line at that wavelength using the values from the simulations described above.  By doing so, we account for incompleteness due to the noise level in the data (from a combination of telescope+instrument throughput, and also integration time), and also due to the night sky lines which are ubiquitous at these wavelengths. 

We perform this emission line simulation above for our GOODS-N observations, measuring the posterior distribution of the expected number of detections as a function of S/N for $e$-folding scales of $W_0=5-200$\AA.  For each choice of $W_0$, we carry out 1000 Monte Carlo simulations.  We illustrate the results for one value of $W_0$ (100 \AA) in the left panel of Figure \ref{fig:n_detection}, which shows the measured number of detections from 100 of the 1000 simulations, highlighting the dispersion in expected number, necessitating the need for a large number of simulations to robustly measure the posterior distribution.  The right panel of Figure \ref{fig:n_detection} displays the mean expected number of detections, averaged over each set of 1000 simulations, as a function of S/N for a range of EW distributions for $6.0 < z < 7.0$.  As seen in this figure, a larger choice of $W_0$ understandably predicts a larger number of detections, as we expect the galaxies to show stronger \lya emission on average.  One strong advantage of this method is that, in addition to our detected \lya emission lines, the large number of non-detections are highly constraining as well. 

\begin{figure*}[t]
\includegraphics[width=1.05\columnwidth]{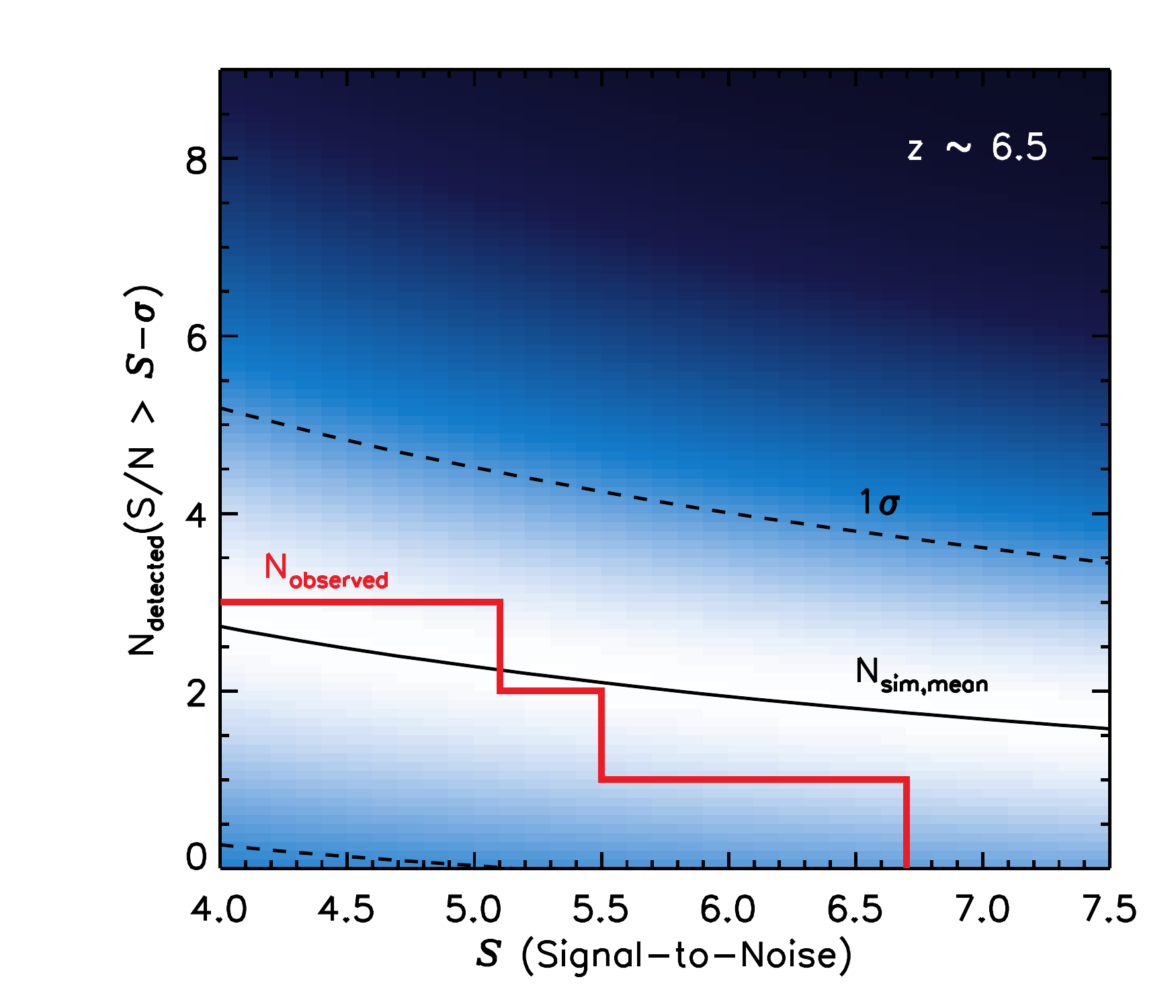}
\includegraphics[width=1.05\columnwidth]{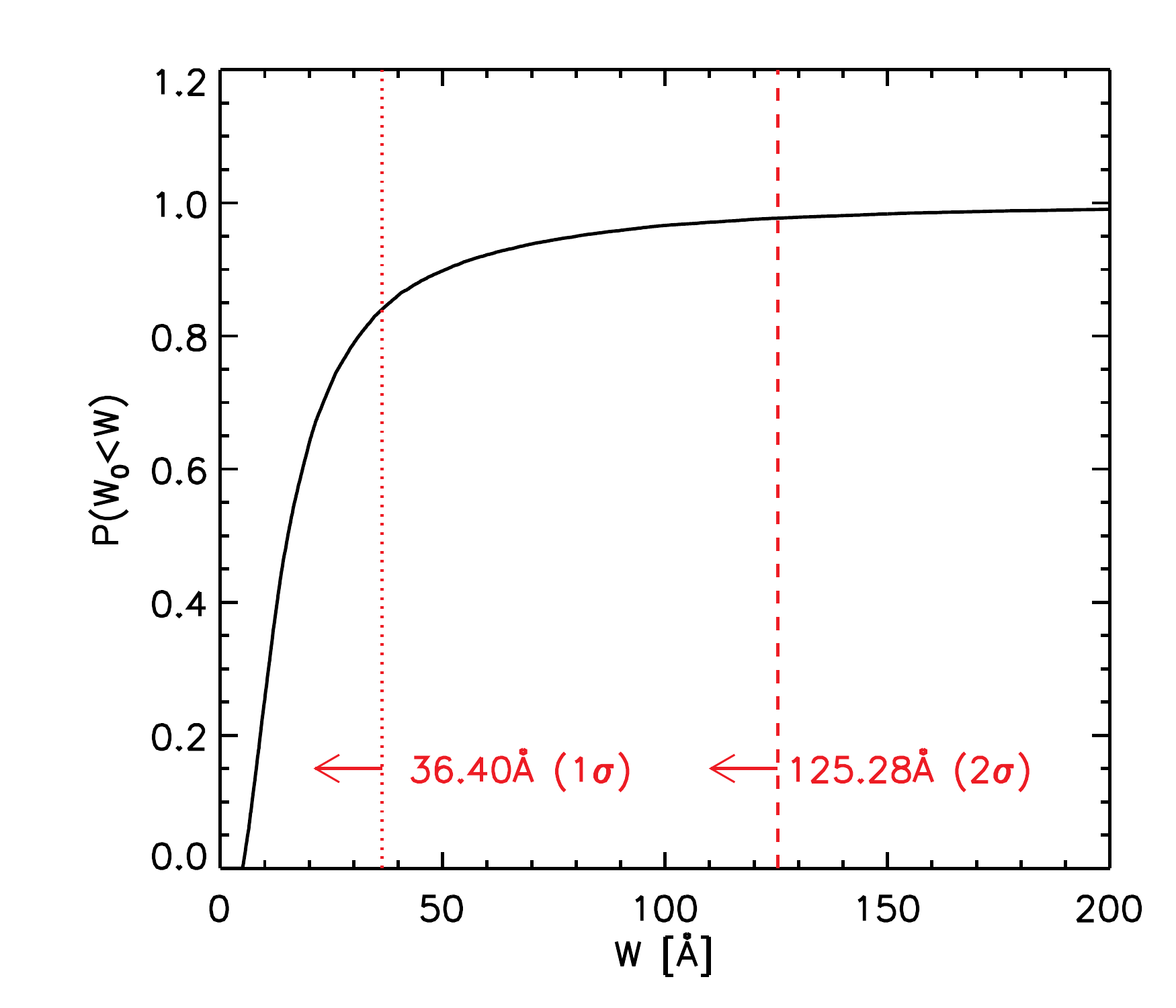}
\caption{(Left) the probability distribution of the expected number of \lya detections as a function of a S/N level ($\mathcal{S}$) at $z\sim6.5$, which is obtained from the $10^5$ MCMC chain steps.  Higher probability regions are denoted by the brighter colors.  The black solid curve shows the mean of the expected number of detections from our simulations as a function of S/N ($\mathcal{S}$), and the dashed curves are 1$\sigma$ uncertainties.  Our three detections are drawn as the red solid line. (Right) the cumulative probability of the EW $e$-folding scale ($W_{0}$) from our MCMC-based fitting algorithm.  The 1$\sigma$ and 2$\sigma$ upper limits are denoted with dotted and dashed red vertical lines, respectively.}
\label{fig:w0_fit}
\end{figure*}

\subsection{An $e$-folding scale of \lya EW Distribution at $z\sim6.5$}
As counting the number of \lya line detections can be described as a general Poisson problem, the likelihood of obtaining the particular results (counting the number of \lya detections) is the Poisson likelihood. A well-known statistic related to the Poisson likelihood is the ``Cash statistic" \citep{Cash1979a}, which is described as follows. 
\begin{eqnarray}
\begin{split}
C &= -2\ ln\mathcal{L}\\
&= -2\sum_{i=1} ({N_{o,i} ln(N_{m,i}) - N_{m,i} - ln N_{m,i}!}),
\end{split}
\label{eqn:likelihood}
\end{eqnarray}
where $\mathcal{L}$ is the Poisson likelihood, $N_{\text{o,i}}$ and $N_{\text{m,i}}$ are the observed and expected number of detections in a corresponding S/N bin, $i$, and $C$ is the goodnees-of-fit statistic so that the expected number of detections matches the observed number of detections in all S/N bins.  $N_{\text{m,i}}$ is calculated based on the choice of $W_0$ as described in the previous section (refer the left panel of Figure \ref{fig:n_detection}).  To construct the probability distribution of $W_0$ with the goodness-of-fit, we carry out MCMC sampling, which uses a Metropolis-Hastings algorithm \citep{Metropolis1953a,Hastings1970a}.  In each chain step, a new candidate value for $W_0$ is randomly drawn from a Gaussian distribution, and we calculate the Poisson distribution log-likelihood of the candidate to go through the acceptance-rejection step.  If the log-likelihood of the candidate $W_0$ exceeds that of the previous one by more than a uniform random variate drawn between 0 and 1, the candidate is accepted and recorded.  Otherwise, the candidate is thrown away and retaken by the previous step.  The random Gaussian width for choosing a new candidate $W_0$ in each step is tuned to have an optimal acceptance rate of 23.4\% \citep{Roberts1997a} to achieve the maximum efficiency of our MCMC sampling.  Before recording the MCMC chains, we also run a burn-in stage to check the convergence of the MCMC sampling.  We employ the Geweke diagnositc \citep{Geweke1992a}, comparing the mean and the variance of the first 10\% of chain steps to those of the last half of samples.  Once the convergence criteria are satisfied, we record the MCMC chains.  After the burn-in stage, we generate $10^5$ MCMC chains, which fully sample the $W_0$ probability distribution.

Performing this MCMC sampling with the three detections (S/N = 5.1, 5.5, and 6.7, respectively) from our observations, we calculate the posterior distribution of the \lya EW $e$-folding scale at $6<z<7$.  Our low number of detections make us unable to robustly constrain the median of $W_0$, thus we find a 1$\sigma$ (84\%) upper limit of $W_{0}<36.4$\AA\ (125.28\AA\ for a 2$\sigma$ limit; see the right panel of Figure \ref{fig:w0_fit}).  In the left panel of Figure \ref{fig:w0_fit}, the background colors represent the probability of the expected number of detections obtained from the $10^5$ MCMC chain steps; higher probability region is denoted by brighter color.  Black solid and dashed curves show the mean and 1$\sigma$ errors of the expected number of detections, and our observational data are shown as a red solid line. 

%%%%%Section5: Discussion%%%%%
\section{Discussion}
\begin{figure*}[t]
\centering
\includegraphics[width=0.65\paperwidth]{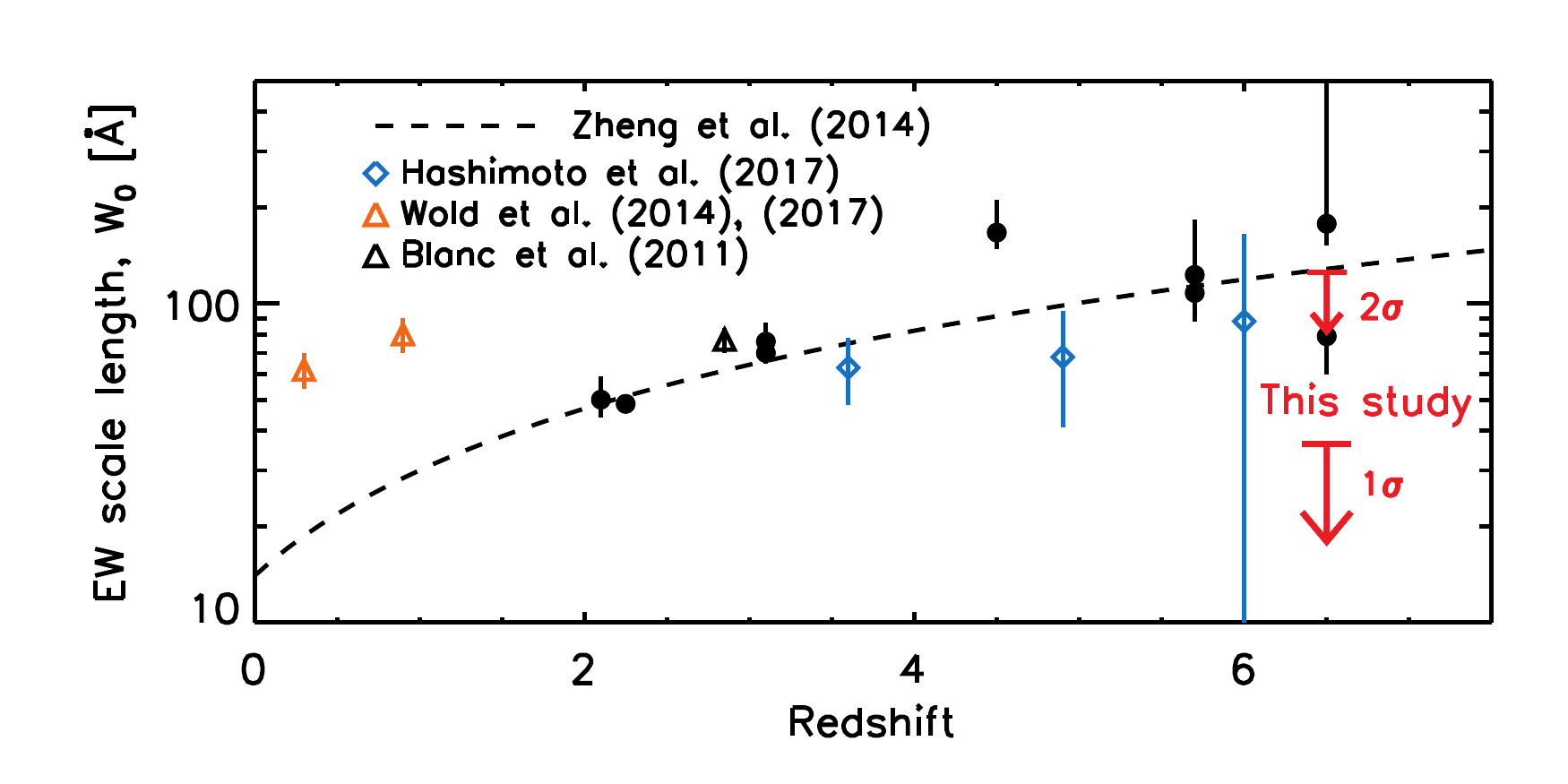}
\caption{The redshift dependence of the \lya EW $e$-folding scale ($W_{0}$).  All data are shown without an IGM absorption correction.  The black dashed line describes the best-fit redshift evolution from \cite{Zheng2014a}, compiling $0<z<7$ LAEs from literature:  \cite{Guaita2010a} at $z=2.1$, \cite{Nilsson2009a} at $z=2.25$, \cite{Gronwall2007a} at $z=3.1$, \cite{Ciardullo2012a} at $z=3.1$, \cite{Ouchi2008a} at $z=3.1, 3.7$, \cite{Zheng2014a} at $z=4.5$, \cite{Kashikawa2011a} at $z=5.7, 6.5$, and \cite{Hu2010a} at $z=5.7, 6.5$ shown as filled circles.  Blue diamonds are the measurements of \cite{Hashimoto2017a} using the LAEs ($M_{UV}<-18.5$) from the MUSE HUDF Survey \citep{Bacon2017a}, which are consistent with \cite{Zheng2014a} at that redshift range.  At lower redshift, the $W_0$ measurements of \cite{Wold2017a} at $z\sim0.3$ and \cite{Wold2014a} at $z\sim0.9$ (orange triangles) suggest a relatively unevolving EW $e$-folding scale of \lya across $z\sim0.3 - 3.0$, considering the other measurements described above, including \citet[black triangle]{Blanc2011a} at $z\sim2.85$.} 
\label{fig:ew_evolution}
\end{figure*}

\subsection{Redshift dependence of the \lya EW $e$-folding scale}
As discussed in the literature, the \lya EW $e$-folding scale, $W_0$, is expected to decrease with increasing neutral hydrogen in the IGM.  \citet[Z14 hereafter]{Zheng2014a} evaluate the redshift dependence of $W_0$ from compiled data at $0<z<7$, which show that larger EW LAEs are found at higher redshift.  More recently, \cite{Hashimoto2017a} report their measurements of $W_0$ at $3<z<6$ using the MUSE HUDF Survey \citep{Bacon2017a}, which are consistent with Z14.

In Figure \ref{fig:ew_evolution}, we compare the redshift dependence of the \lya EW $e$-folding scale from previous studies to our measure.  Compared to the derived evolution of Z14 (black dashed curve), our measurement (red arrows) shows that, at 1$\sigma$ confidence, this quantity must begin to drop at $z >$ 6. As we expect that a higher fraction of neutral hydrogen in the IGM would reduce the strength of \lya emission and lower the EW $e$-folding scale, this drop can be interpreted as a signal of an increasing amount of neutral hydrogen in the IGM, although the literature measurements of $W_0$ at $4 < z < 6$ are consistent with our measure at the 2$\sigma$ level.  

However, the recent study of \cite{De-Barros2017a} presents a lower \lya fraction at $z\sim6$ than the values previously reported in the literature.  Although they measure the Ly$\alpha$ fraciton and not the EW distribution, this could mitigate the tension between our 1$\sigma$ upper limit and previous results, implying perhaps no significant evolution from $z =$ 6 to 6.5, though increasing the confidence of significant evolution at $z =$ 4 to 6.  This is confirmed by \cite{Mason2017a}, who parameterize the $z\sim6$ \lya EW distribution of \cite{De-Barros2017a} as a function of $M_{\text{UV}}$, Eq. (4) in their paper, and find an $e$-folding scale of the $z\sim6$ \lya EW distribution from their parameterization ranges from $W_0 =$19 -- 43\AA\ (with $M_{\text{UV}}=$-17.5 to -22.5), also lower than those from  other studies in the literature at $4 < z < 6$.  With the drop of the EW $e$-folding scale at $z\sim6.5$ from our measurement, the recent measurements of the EW $e$-folding scale imply the smoother evolution of the neutral hydrogen fraction in the IGM between $z < 6$ and $z > 6$.

Also, it is worth mentioning the known effect that UV-selected LAEs have larger EWs with fainter UV magnitudes \citep[e.g.,][]{Ando2006a, Stark2010a, Stark2011a, Schaerer2011a, Cassata2015a, Furusawa2016a, Wold2017a}. \cite{Hashimoto2017a} systematically test the effect of sample selection on measuring the \lya EW $e$-folding scale and find that including UV-fainter LAEs increases the measured $e$-folding scale of the \lya EW distribution (refer to Figure 8 in their paper).  Shown in Figure \ref{fig:muv_z}, our photo-$z$ selected galaxies have UV magnitudes $\lesssim-18.5$ (in GOODS-N), missing very UV-faint galaxies.  Therefore, our measure of the EW $e$-folding scale can be biased toward a small value.  However, the drop at $z\sim6.5$ of the measured EW $e$-folding scale found in this study is not fully explained by the sample selection effect and is still significant, compared to those at lower redshifts which use the similar UV magnitude cut ($M_{\text{UV}}<-18.5$) \citep[e.g.,][]{Hashimoto2017a}.  Incorporating additional data from our MOSFIRE observations in our follow-up paper will update this result and its statistical confidence, and for further constraints, a more comprehensive analysis accounting for the UV magnitude dependence of the \lya strength is needed in future study. 

\subsection{Testing our measure of the \lya EW $e$-folding scale}
\begin{figure}[t]
\includegraphics[width=\columnwidth]{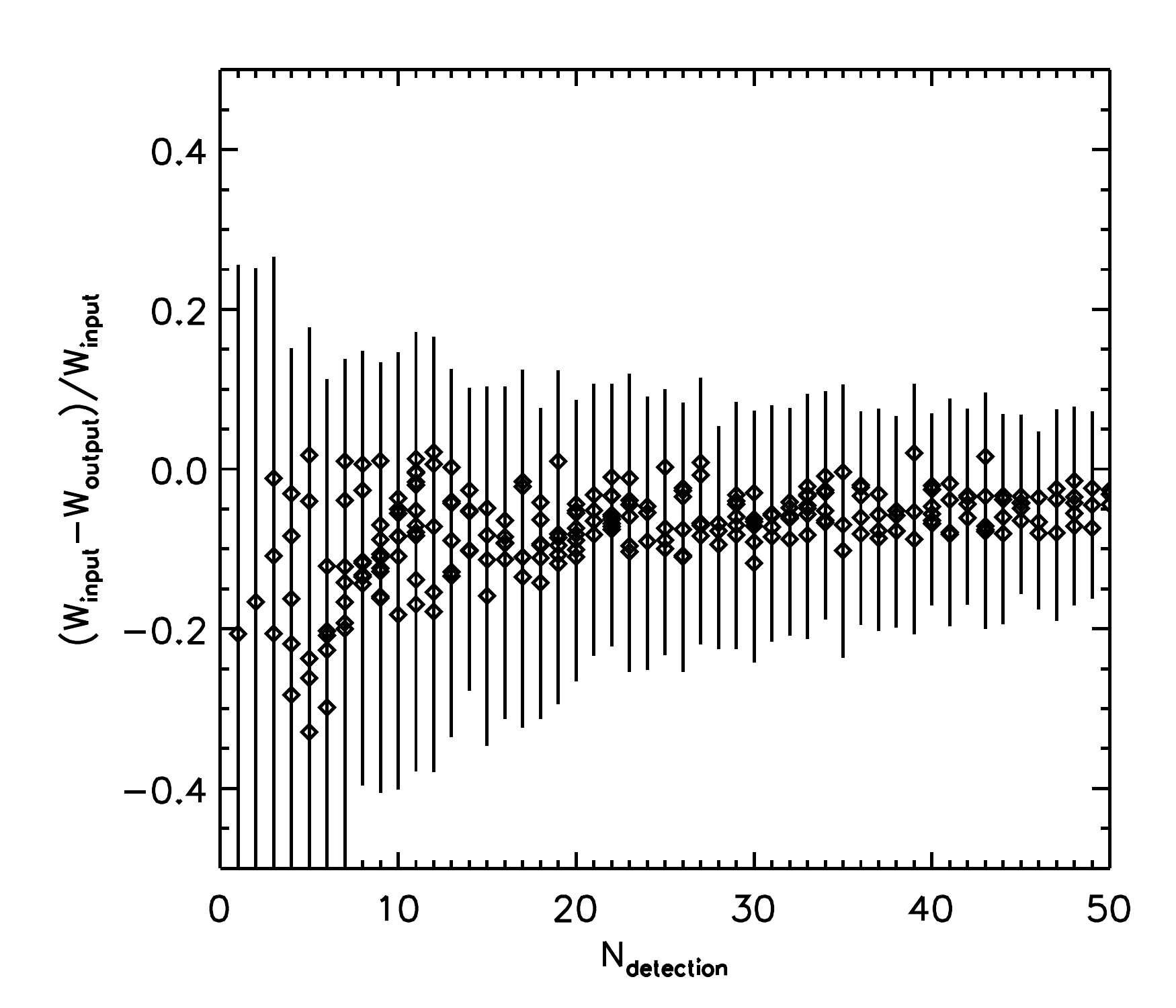}
\caption{Fractional error of the recovered $e$-folding scale compared to the input as a function of the number of mock detections used in our fitting procedure.  We fit the $e$-folding scale with mock detections which follow the assumed EW distributions with the $e$-folding scale ranging from 5 to 200\AA.  For each assumed $e$-folding scale, we create 1000 sets of mock detections in a Monte-Carlo fashion and recover the $e$-folding scale.  The medians of the fractional errors of the $e$-folding scale are shown as diamonds with the error bars denoting the standard deviation.  With $\lesssim10$ detections, our simulation recovers the true $e$-folding scale within $\lesssim30\%$, though there is a bias towards recovering more galaxies than those input due to up-scattering by noise near to the detection limit.  However, an increased number of detections mitigate this bias, ensuring a more accurate recovery of the true $e$-folding scale. }
\label{fig:w0_test}
\end{figure}

We provide a measure of the \lya EW $e$-folding scale at the end of reionization using our comprehensive simulations for predicting the expected number of \lya detections.  A novel way of accounting for the data incompleteness allows us to constrain the observed EW distribution of \lya lines with a handful of detections as all non-detections are highly constraining in our simulations.  This is very promising for upcoming additional spectroscopic studies of high-redshift \lya emitting galaxies.  Here we test the ability of our simulation to recover the \lya EW $e$-folding scale using sets of virtual observations, and show how much future spectroscopic searches can improve the constraining power of measuring the EW $e$-folding scale. 

To do the recovery test, we create sets of virtual observations. In each set of virtual observations, we first generate mock emission lines with $N_{\text{detection}} = 1$ -- 50, following the likelihood of the expected detections as a function of S/N level, which is derived from our simulation in Section 4.1.  With the virtual set of detections, we fit the EW $e$-folding scale as described in Section 4.2.  For each set of \textit{virtually} detected emission lines ($N_{\text{detection}}=1-50$), we create 1000 sets of virtual observations in a Monte-Carlo fashion, and recover the EW $e$-folding scale from each virtual dataset in the same was as done on our real data.  Figure \ref{fig:w0_test} shows median and standard deviation of the fractional error of the recovered $e$-folding scale to the input as a function of the number of virtual detections used in our fitting procedure.  With $\lesssim10$ detections, our simulation recovers the true $e$-folding scale with $\lesssim30\%$ of fractional errors.  This does show a systematic bias on the derived EW $e$-folding scale, in that the recovered values are often lower than the true values with small numbers ($\lesssim10$) of detections.  This bias is lessened with increasing numbers of detections, and the spread in this fractional error also decreases.  One cause of this bias could be that in the limit where few lines are seen, these lines are likely close to the detection limit, where noise fluctuations can up-scatter lines below the limit to the detectable level, in a form of Eddington bias.  However, an increasing number of mock detections mitigate this bias, ensuring the recovery of the assumed $e$-folding scale with smaller biases ($\lesssim10\%$ with $\gtrsim20$ detections) and smaller errors.  Due to our small number of detected lines ($N_{\text{detection}}=3$), our measure of the \lya EW $e$-folding scale at $z\sim6.5$ may be subject to this bias.  However, the direction of this bias would cause us to overestimate the number of detected lines, resulting in our quoted upper limits on $W_0$ being conservatively high.

\subsection{\lya Detection Probability as a Function of Redshift}

\begin{figure}[t]
\centering
\includegraphics[width=\columnwidth]{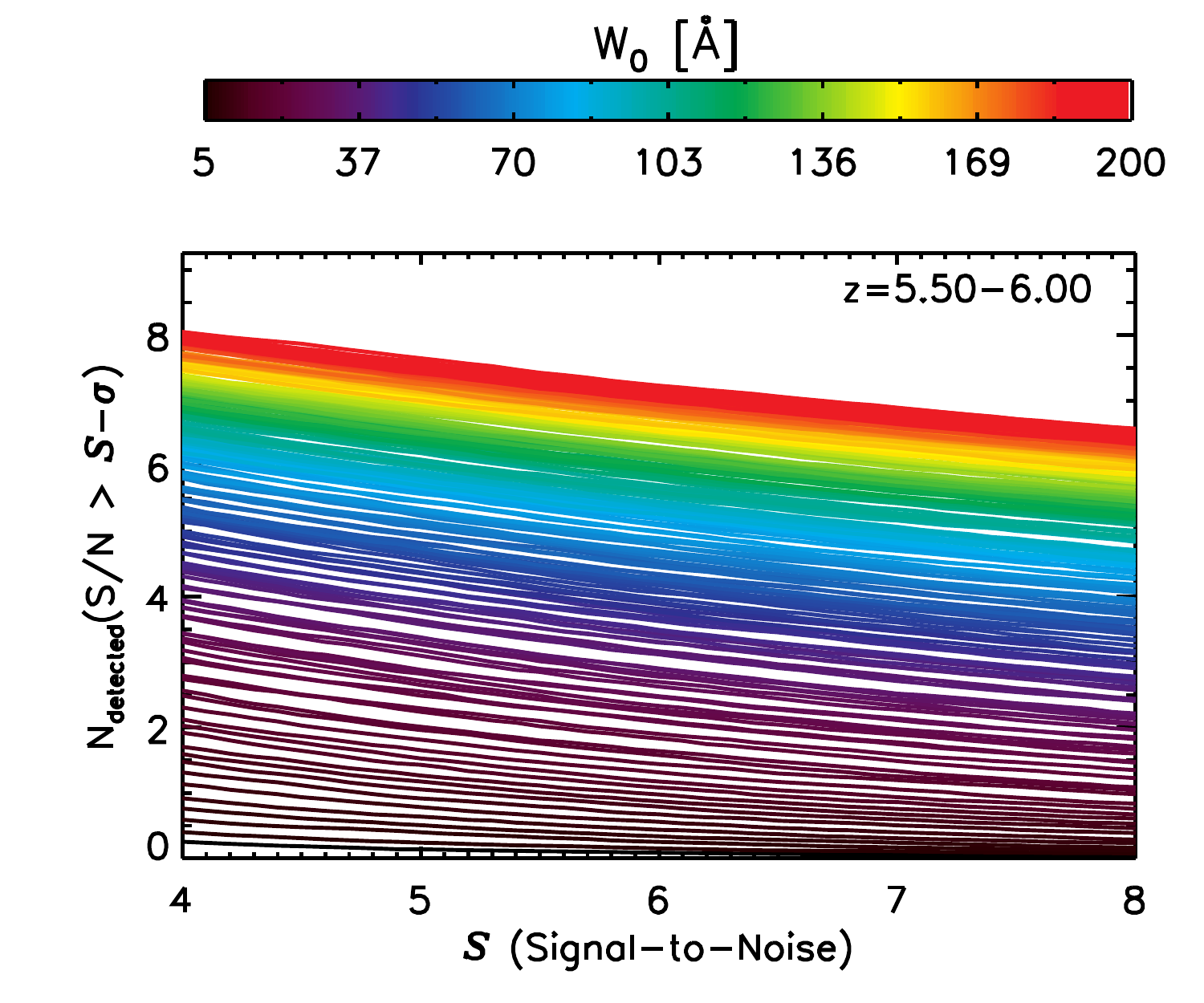}
\caption{The expected number of detections as a function of S/N level ($\mathcal{S}$) with various EW distributions at $z=5.5-6.0$.  Unlike the predicted number of detections, we do not detect any \lya emission in this range.  A large spectroscopic survey with \lya detections is needed to constrain the \lya EW distribution.}
\label{fig:n_detection_z55}
\end{figure}

\begin{figure}[t]
\centering
\includegraphics[width=\columnwidth]{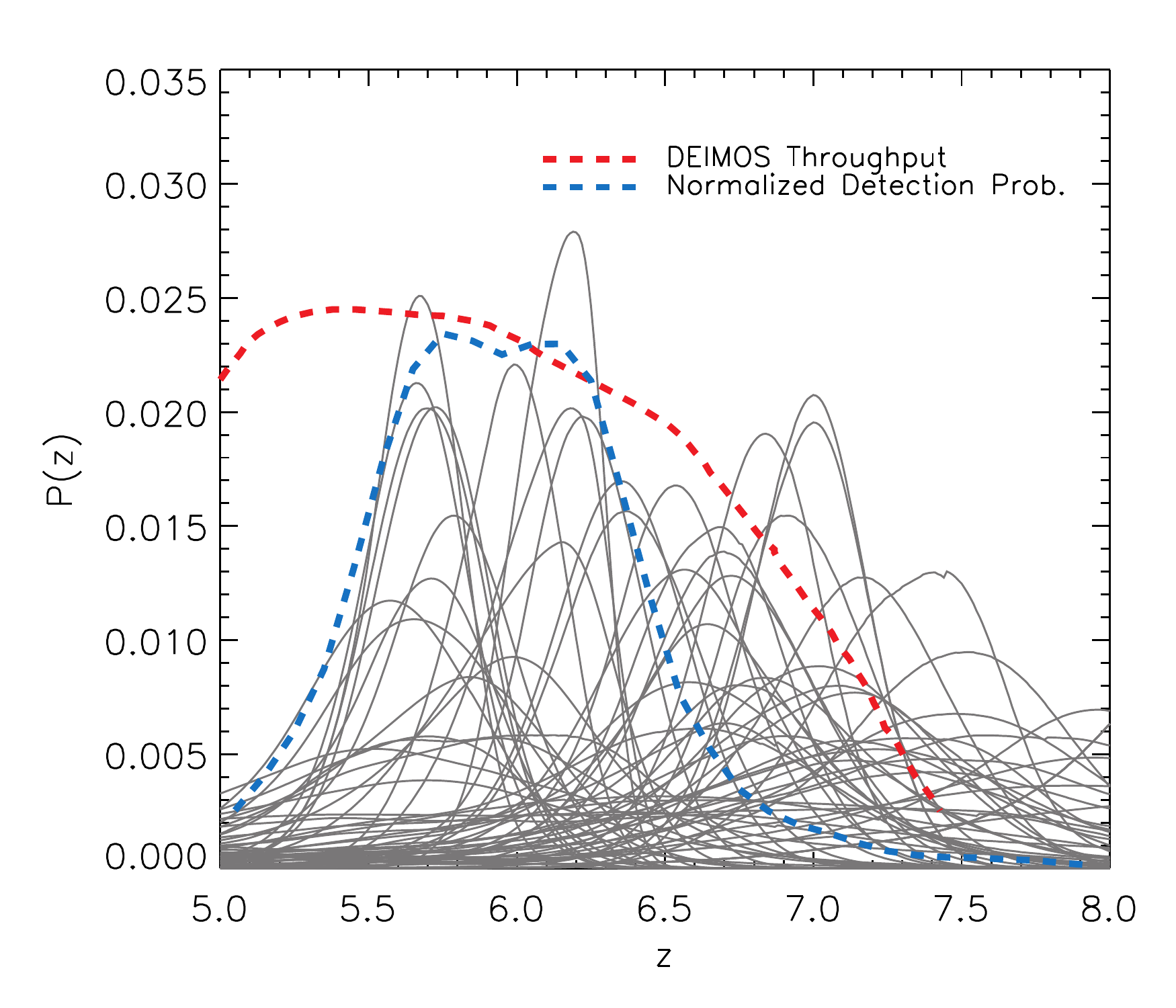}
\caption{\lya detection probability (blue dashed curve) as a function of redshift in our spectroscopic dataset. The probability is measured, accounting for the probability distribution functions, $P(z)$, of the photometric redshifts of our target galaxies (grey curves) and the DEIMOS instrument throughput (red dashed curve). The estimated \lya detection probability is high at $z\sim5.5-6.5$, and declines at $z\gtrsim7$.}
\label{fig:pz_dist}
\end{figure}

We have no \lya detections at $5.5 < z < 6.0$ in our GOODS-N data, although the DEIMOS observations are sensitive to \lya in that redshift range.  As shown in Figure \ref{fig:n_detection_z55}, our simulations estimate that at $5.5 < z < 6.0$ we should have detected at least a couple of \lya lines, as many as eight detections with a large $e$-folding scale of $W_0=200$\AA.  At the published values at $5.5 < z < 6.0$ of $W_0\sim100$\AA\ (see Figure \ref{fig:ew_evolution}) our simulations predict $N_{\text{detection}}=6.62\pm1.99$.  In Figure \ref{fig:pz_dist} we show the galaxy photometric redshift probability distribution functions for our sample (grey curves) and the DEIMOS instrument throughput (red dashed curve).  Combining the two we calculate the \lya detection probability (blue dashed curve), which is the normalized-expected number of \lya detections among our target galaxies as a function of redshift accounting for this throughput, which is high at $z\sim5.5-6.5$, and declines at $z\gtrsim7$.  This shows that non-detections are understandable at $z\sim7$, but the lack of detections at $z<6$ is somewhat unexpected.  This could be due to the inhomogeneous nature of cosmic reionization at the very end, but it also reiterates the need for a more comprehensive spectroscopic survey over larger area to marginalize over these small number statistics.

\subsection{Systematic Effects of Photometric Redshifts}
\begin{figure}[t]
\centering
\includegraphics[width=\columnwidth]{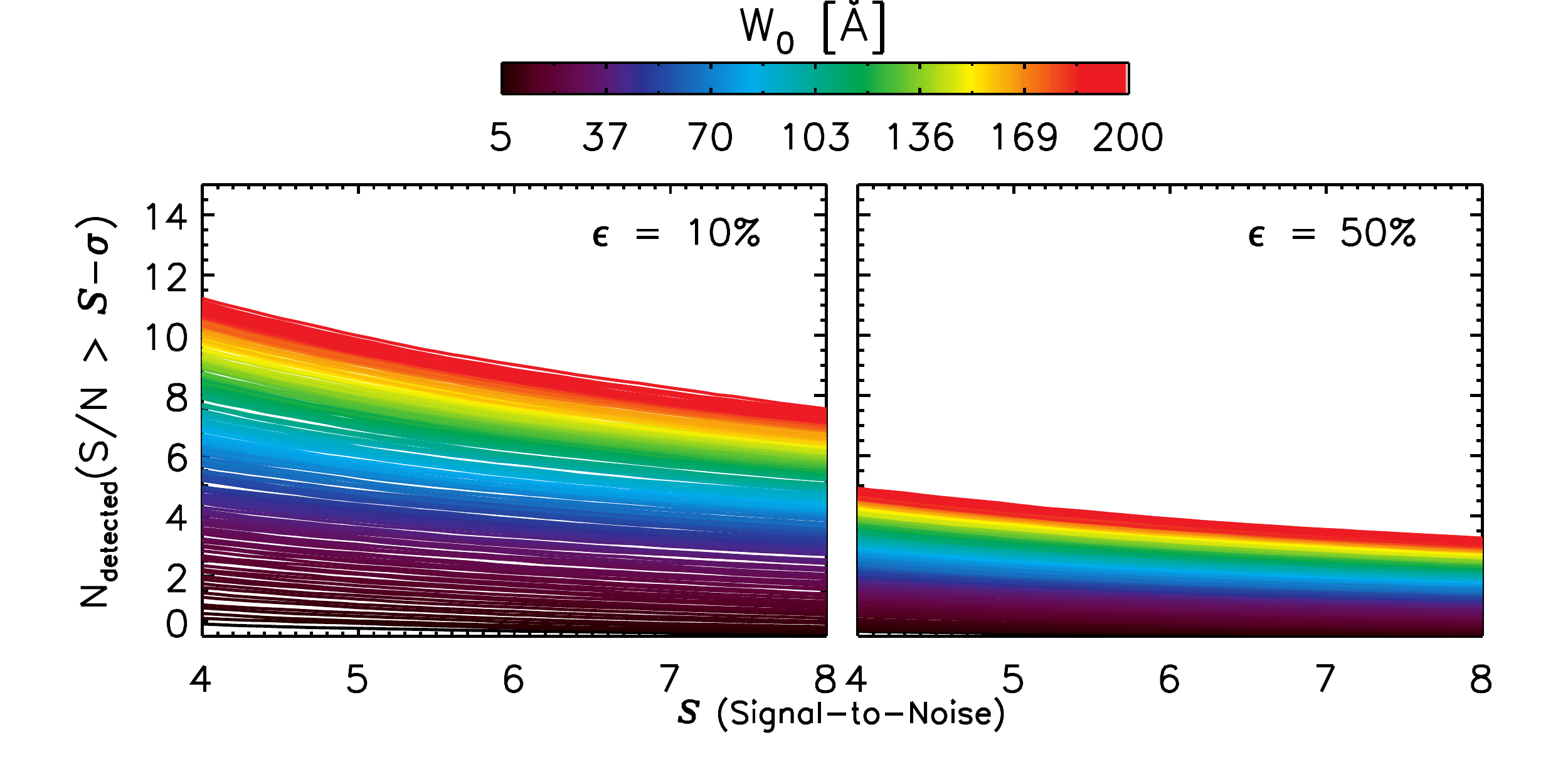}
\centering
\includegraphics[width=\columnwidth]{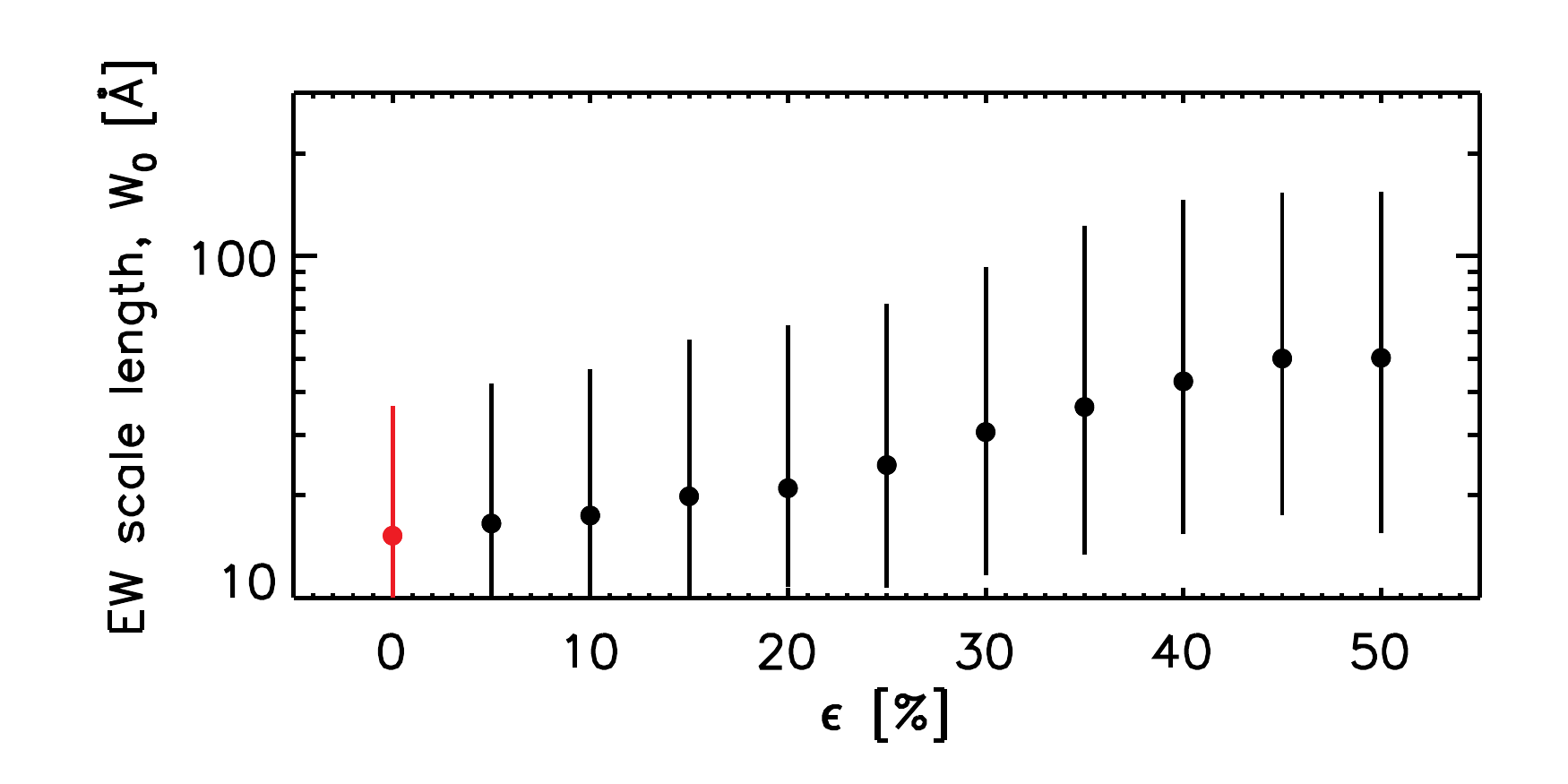}
\caption{Top: the expected number of detections measured as described in Section 4.1, but when increasing the photometric redshift uncertainties by 10\% (left) and 50\% (right).  Increasing the $P(z)$ uncertainty reduces $N_{\text{detected}}$ as the \lya emission lines have more chance to be found outside of the instrumental wavelength coverage; the predicted $N_{\text{detected}}$ is almost half the number of our fiducial result (Figure \ref{fig:n_detection}) in the case that $P(z)$ has an additional 50\% systematic error.  Bottom: The \lya EW $e$-folding scale as a function of the augmented systematic errors in $P(z)$.  As the augmented systematic error on $P(z)$ increases, the fitted \lya EW $e$-folding scale in our simulation increases as well.  If the true uncertainties were $\gtrsim$50\% of those assumed, our observations could be consistent with no evolution in the Ly$\alpha$ EW distribution.}
\label{fig:pz_test}
\end{figure}

As spectroscopic confirmation of galaxies at $z\gtrsim6$ is observationally expensive, photometric redshift selection provides a powerful means to construct extensive high-redshift galaxy catalogs based on multi-wavelength imaging survey data \citep[e.g.,][]{Stark2009a, Papovich2011a, Dunlop2013a, McLure2013a, Schenker2013a, Bouwens2015a, Finkelstein2015a, Salmon2015a, Song2016a, Livermore2017a}.  However, photometric redshift measurements alone cannot completely remove the possibility that high-redshift candidate galaxies can be low-redshift interlopers.  More pressingly, a comprehensive analysis on the accuracy of the photometric redshift measurements, specifically a calibration of these PDFs, at $z\gtrsim6$ is lacking.  Thus statistical studies using photometric redshifts could be biased if there are uncovered systematic effects in photometric redshifts.  Interestingly, a recent spectroscopic confirmation of \lya emission from the $HST$ Faint Infrared Grism Survey (FIGS) \citep[PI: S. Malhotra;][]{Pirzkal2017a} finds a \lya emission line at $z=$7.452 which is different from the photometric redshift at the 2$\sigma$ level \citep{Larson2017a}.  While these differences should happen a small fraction of the time, here we consider the effects of underestimating the photometric redshift PDF.

To test how accurate our measure of the \lya EW $e$-folding scale is to an increased photometric redshift uncertainty, we smooth the probability distribution functions of photometric redshifts to simulate a potential systematic underestimation of the errors of the current photometric redshift measurement, and perform our simulations on this altered dataset.  The top panels of Figure \ref{fig:pz_test} show the likelihood of detecting \lya emission lines with errors augmented by 10\% (left) and 50\% (right), which predict fewer detections than that with the current errors on the photometric redshifts.  If the errors on the photometric redshifts are underestimated, our simulations predict too many \lya detections, and our measured EW $e$-folding scale could be biased toward smaller values.  A systematic test of the increased errors of photometric redshifts is shown in the bottom panel of Figure \ref{fig:pz_test}.  If the true uncertainty of photometric redshift is 50\% larger than the current estimate, our constrained EW $e$-folding scale would be increased by a factor of a few: $W_0<154.68$\AA\ at 1$\sigma$ confidence.  If this is the case, then our lack of detections is consistent with \emph{no} evolution in the Ly$\alpha$ EW distribution at $z >$ 6, highlighting the importance of calibrating the photometric redshift uncertainties with a dedicated spectroscopic survey, likely to come with the advent of the {\it James Webb Space Telescope}.

%%%%%Section6: Summary%%%%%
\section{Summary}
We have collected four nights of spectroscopic observations over 118 galaxies at $z\sim5-7$ in the GOODS fields with DEIMOS on the Keck telescope to search for \lya emission in the early universe.   We use these data to provide a new constraint for the $e$-folding scale of the \lya EW distribution at the end of reionization.  We simulate the predicted number of \lya detections at a given expected S/N value in our observational data with a range of \lya EW distributions (parameterized by the $e$-folding scale, $W_0$).  We comprehensively account for incompleteness due to observational conditions (e.g., integration time, sky emission, and instrument throughput) as well as galaxy photometric redshift probability distribution functions.  With our three detected Ly$\alpha$ lines in the GOODS-N field, we constrain the characteristic $e$-folding scale of the \lya EW distribution at $z\sim6.5$.  Our main results are summarized as follows. 
\begin{enumerate}
\item Performing an automated search for emission lines in 1D spectra as well as visual inspection of 2D spectra, we detect five emission lines above a 5$\sigma$ significance level from four nights of Keck/DEIMOS observation among a sample of 118 high-$z$ candidate galaxies in the GOODS-S and GOODS-N fields.  Our tests of the possibility of low-$z$ interlopers indicates that the detected lines are likely \lya emission at $z\gtrsim5.5$. 
\item We simulate the expected number of \lya detections from our observational dataset, comprehensively taking into account noise in the dataset and galaxy photometric redshift probability distributions.  In the simulations, we construct the probability distribution of the expected number of detections as a function of S/N level with various $e$-folding scales ($W_0$) of the \lya EW distribution, where a larger value of $W_0$ predicts a larger number of Ly$\alpha$ detections.
\item Our dataset constrains the \lya EW $e$-folding scale at $z\sim6.5$ to be $<$ 36.40\AA\ at 1$\sigma$ confidence (125.28\AA\ at 2$\sigma$).  This is lower than previous measurements at lower redshifts, providing weak evidence for an increasing fraction of neutral hydrogen in the IGM at this epoch.  Additional data from our MOSFIRE observations at $z>7$ will update this result in a future paper with a higher statistical confidence.
\item We test the ability of our simulation to recover the \lya EW $e$-folding scale as a function of the number of detections, and find $\gtrsim20$ detections allow us to recover the true value of the \lya EW $e$-folding scale to $\lesssim10\%$ accuracy; these simulations imply that our current results provide conservative upper limits.  We also find that systematic errors in the photometric redshift uncertainties would have a significant impact on constraining the EW $e$-folding scale, suggesting that a comprehensive analysis of photometric redshift uncertainties in the early universe is necessary. 
\end{enumerate}

As mentioned in Section 2.1 and shown in Figure \ref{fig:mask_design}, in addition to DEIMOS, our entire spectroscopic dataset utilizes MOSFIRE as well to search \lya emission at $z>7$, and we will analyze these MOSFIRE observations in a follow-up paper, using the DEIMOS observations in this paper as the $z =$ 6.5 anchor.  Furthermore, analyzing both the DEIMOS and MOSFIRE data allows us to search for \lya emission more comprehensively, which guarantees to cover the entire wavelength range of the probable locations of \lya emission from galaxies at $5<z<8.2$. Particularly for $z\sim7$ candidate galaxies, \lya emission from these galaxies cannot be entirely searched by a single instrument, either DEIMOS or MOSFIRE (which observes \lya emission at $5<z<7$ and $7<z<8.2$, respectively).  Thus, a complete search for \lya emission at $z\sim7$ with a systematic spectroscopic survey with both DEIMOS and MOSFIRE will make significant progress.

Our measurement of the \lya EW distribution alone is not enough to calculate directly the IGM neutral hydrogen fraction during reionization.  We require detailed models, which account for a variety of effects, to calculate a likely range for the neutral fraction based on our observations \cite[see discussion in][]{Bolton2013a, Dijkstra2014a}. Combined with these models our \lya visibility can constrain the neutral fraction of hydrogen in the IGM, but this constraining power depends on several factors, which need to be considered in future studies: the amount of residual neutral hydrogen in the circumgalactic medium, the number of self-shielding Lyman-limit systems in the IGM, where \lya is self-shielded by overdense gas, and the Lyman continuum escape fraction.  By providing our estimate of the \lya EW distribution at $z>6$, we provide observational constraints to reionization models, leading to improved predictions during the end of reionization over $6 < z < 8$.

\acknowledgments
The authors wish to recognize and acknowledge the very significant cultural role and reverence that the summit of Mauna Kea has always had within the indigenous Hawaiian community.  We are most fortunate to have the opportunity to conduct observations from this mountain.  We also thank M. Dijkstra for constructive discussions.  I.J. acknowledges support from the NASA Headquarters under the NASA Earth and Space Science Fellowship Program - Grant 80NSSC17K0532.  I.J. and S.F. acknowledge support from NSF AAG award AST-1518183.  RCL acknowledges support from an Australian Research Council Discovery Early Career Researcher Award (DE180101240).  M.S.'s research was supported by an appointment to the NASA Postdoctoral Program at the NASA Goddard Space Flight Center, administered by  Universities Space Research Association under contact with NASA.

%%%%%Appendix: Flux calibration%%%%%
\appendix
\section{Flux calibration}
We calculated the flux calibration for the GOODS-N observations to be accurate at the $\sim20\%$ accuracy level, but the calibrated flux in the GOODS-S observations has a much larger random error.  Comparing the fluxes of point sources derived from our DEIMOS data to those from $HST$ imaging \citep{Finkelstein2015a}, we find large deviations at a level of $>$100\% (see Column 5 and 6 in Table \ref{tab:calibration}).  With this level of flux uncertainty, we are unable to constrain the \lya EW for sources in the GOODS-S mask, so we exclude the GOODS-S data in our analysis of estimating the EW $e$-folding scale, though we note the two new spectroscopic redshifts from these data are robust.  We discuss the issues with flux calibration of the GOODS-S observations in more detail here.

We have standard star observations with a long slit (1.0$\arcsec$ slit width), the same as that of the science observations.  To calibrate fluxes precisely, we take into account the possible slit losses with respect to the measured seeing for science objects and standard stars.  However, we lack continuum sources observed with the 1.0$\arcsec$-width slit in the science frames, which are necessary to provide a cross-check on our flux calibration.  Instead, the only available continuum sources in the science frames are guide stars observed in 4.0$\arcsec$$\times$4.0$\arcsec$ square-box slits.  Thus, we calculate continuum fluxes of guide stars in the science frames to check our flux calibration, accounting for the differences in slit losses due to the larger boxes, and compare those to the $HST$ imaging magnitudes from \cite{Finkelstein2015a}.  The derived continuum fluxes with the $HST$/ACS 850LP ($z_{850}$-band) filter curve are listed in the last column of Table \ref{tab:calibration}.

Another issue is that we dither our observations with a $1.0\arcsec$ drift.  Using dithered data, it is critical to locate spatial positions of the objects in tge 2D spectra (centering in y-direction).  However, our standard star (G191-B2B) observation in GOODS-S shows large drifts on centering (up to $\sim5$ pixels in the spatial direction), compared to that ($\lesssim1$ pixel) of the GOODS-N standard star observations (BD$+$33d2642).  Even though we account for a possible atmospheric dispersion in our data reduction as described in Section 2.2 (which is successful for the GOODS-N flux calibration as shown in the top four rows in Table \ref{tab:calibration}), the large drift of centering 2D spectra of the standard star observations in GOODS-S results in significant uncertainties (the bottom three rows in the same table). 

\begin{table}[h]
\centering
\begin{center}
\caption{Summary of the $z_{850}$ Magnitudes of Point Sources}
\begin{tabular}{cccccc}
\tableline 
\tableline
\quad {Field} & {Source ID} 	    & {R.A.}         & {Decl.}        & {$z_{850}$}	   		 & {$z_{850}$}  		 \\
\quad {}         & {}  			    & {(J2000.0)} & {(J2000.0)} & {\cite{Finkelstein2015a}}   & {DEIMOS\tablenotemark{a}}  \\
\quad {(1)}         & {(2)}  			    & {(3)} & {(4)} & {(5)}   & {(6)}  \\
\tableline
\quad {GOODS-N} &{star\_14984} 	    & {189.1057292} & {62.2346917} & {16.873200} & {16.683006}  \\
\quad {} &{star\_21581} 	    & {189.1264208} & {62.2506417} & {21.465400} & {21.560181}  \\
\quad {} &{star\_33057} 	    & {189.2163417 } & {62.3280833} & {18.386700} & {18.652014}  \\
\quad {} &{star\_34028} 	    & {189.1057292} & {62.2969889} & {17.186700} & {16.954541}  \\
\tableline
\quad {GOODS-S} &{star\_8773} 	 & {53.2407833} & {-27.8863806} & {18.888600} & {17.924650, 18.628263}\\
\quad {}                  &{star\_15723}     & {53.2318250 } & {-27.8572750} & {15.667800} & {14.062555, 14.760222} \\
\quad {}                  &{star\_44065}     & {53.1327917} & {-27.6848694} & {17.030300} & {15.927838, 16.599801} \\
\tableline
\end{tabular}
\end{center}
\begin{flushleft}
\tablecomments{Here we show the calibrated magnitudes of point sources from the first night of observation for GOODS-N and both nights for GOODS-S.  The measure of the $z$-band magnitude in the DEIMOS calibrated stellar spectra (column 6), which accounts for differential slit losses, should equal the known magnitude from the CANDELS catalog (column 5).  From our flux calibration, the calculated flux for the GOODS-N observations is accurate at the $\sim20\%$ accuracy level, while the calibrated flux in the GOODS-S observations has a much larger random error ($>$100\%).}
\end{flushleft}
\label{tab:calibration}
\end{table}


\begin{thebibliography}{}
\footnotesize
\expandafter\ifx\csname natexlab\endcsname\relax\def\natexlab#1{#1}\fi


\bibitem[Ando et al.(2006)]{Ando2006a} Ando, M., Ohta, K., Iwata, I., et 
al.\ 2006, \apjl, 645, L9 


\bibitem[Bacon et al.(2017)]{Bacon2017a} Bacon, R., Conseil, S., Mary, D., 
et al.\ 2017, \aap, 608, A1 


\bibitem[Becker et al.(2001)]{Becker2001a} Becker, R.~H., Fan, X., White, 
R.~L., et al.\ 2001, \aj, 122, 2850 


\bibitem[Blanc et al.(2011)]{Blanc2011a} Blanc, G.~A., Adams, J.~J., 
Gebhardt, K., et al.\ 2011, \apj, 736, 31 


\bibitem[Bolton et al.(2011)]{Bolton2011a} Bolton, J.~S., Haehnelt, M.~G., 
Warren, S.~J., et al.\ 2011, \mnras, 416, L70 


\bibitem[Bolton 
\& Haehnelt(2013)]{Bolton2013a} Bolton, J.~S., \& Haehnelt, M.~G.\ 2013, \mnras, 429, 1695 


\bibitem[Bouwens et al.(2014)]{Bouwens2014a} Bouwens, R.~J., Illingworth, 
G.~D., Oesch, P.~A., et al.\ 2014, \apj, 793, 115 


\bibitem[Bouwens et al.(2015)]{Bouwens2015a} Bouwens, R.~J., Illingworth, 
G.~D., Oesch, P.~A., et al.\ 2015, \apj, 803, 34 


\bibitem[Bruzual 
\& Charlot(2003)]{Bruzual2003a} Bruzual, G., \& Charlot, S.\ 2003, \mnras, 344, 1000 


\bibitem[Caruana et al.(2012)]{Caruana2012a} Caruana, J., Bunker, A.~J., 
Wilkins, S.~M., et al.\ 2012, \mnras, 427, 3055 


\bibitem[Caruana et al.(2014)]{Caruana2014a} Caruana, J., Bunker, A.~J., 
Wilkins, S.~M., et al.\ 2014, \mnras, 443, 2831 


\bibitem[Caruana et al.(2018)]{Caruana2018a} Caruana, J., Wisotzki, L., 
Herenz, E.~C., et al.\ 2018, \mnras, 473, 30 


\bibitem[Cash(1979)]{Cash1979a} Cash, W.\ 1979, \apj, 228, 939 


\bibitem[Cassata et al.(2015)]{Cassata2015a} Cassata, P., Tasca, L.~A.~M., 
Le F{\`e}vre, O., et al.\ 2015, \aap, 573, A24 


\bibitem[Chardin, Puchwein, 
\& Haehnelt(2017)]{Chardin2017a} Chardin, J., Puchwein, E., \& Haehnelt, M.~G.\ 2017, \mnras, 465, 3429 


\bibitem[Ciardullo et al.(2012)]{Ciardullo2012a} Ciardullo, R., Gronwall, 
C., Wolf, C., et al.\ 2012, \apj, 744, 110 


\bibitem[Cooper et al.(2012)]{Cooper2012a} Cooper, M.~C., Newman, J.~A., 
Davis, M., Finkbeiner, D.~P., 
\& Gerke, B.~F.\ 2012, Astrophysics Source Code Library, ascl:1203.003 


\bibitem[Cowie, Barger, 
\& Hu(2011)]{Cowie2011a} Cowie, L.~L., Barger, A.~J., \& Hu, E.~M.\ 2011, \apj, 738, 136 


\bibitem[Cowie, Barger, 
\& Hu(2010)]{Cowie2010a} Cowie, L.~L., Barger, A.~J., \& Hu, E.~M.\ 2010, \apj, 711, 928 


\bibitem[Curtis-Lake et al.(2012)]{Curtis-Lake2012a} Curtis-Lake, E., 
McLure, R.~J., Pearce, H.~J., et al.\ 2012, \mnras, 422, 1425 


\bibitem[D'Aloisio, McQuinn, Davies, 
\& Furlanetto(2018)]{DAloisio2018a} D'Aloisio, A., McQuinn, M., Davies, F.~B., \& Furlanetto, S.~R.\ 2018, \mnras, 473, 560 


\bibitem[D'Aloisio et al.(2017)]{DAloisio2017a} D'Aloisio, A., Upton 
Sanderbeck, P.~R., McQuinn, M., Trac, H., 
\& Shapiro, P.~R.\ 2017, \mnras, 468, 4691 


\bibitem[De Barros et al.(2017)]{De-Barros2017a} De Barros, S., Pentericci, 
L., Vanzella, E., et al.\ 2017, \aap, 608, A123 


\bibitem[Dijkstra(2014)]{Dijkstra2014a} Dijkstra, M.\ 2014, \pasa, 31, e040 


\bibitem[Dijkstra et al.(2014)]{Dijkstra2014b} Dijkstra, M., Wyithe, S., 
Haiman, Z., Mesinger, A., \& Pentericci, L.\ 2014, \mnras, 440, 3309 


\bibitem[Dunlop et al.(2013)]{Dunlop2013a} Dunlop, J.~S., Rogers, A.~B., 
McLure, R.~J., et al.\ 2013, \mnras, 432, 3520 


\bibitem[Fan et al.(2006)]{Fan2006a} Fan, X., Strauss, M.~A., Becker, R.~H., 
et al.\ 2006, \aj, 132, 117 


\bibitem[Finkelstein et al.(2013)]{Finkelstein2013a} Finkelstein, S.~L., 
Papovich, C., Dickinson, M., et al.\ 2013, \nat, 502, 524 


\bibitem[Finkelstein et al.(2011)]{Finkelstein2011a} Finkelstein, S.~L., 
Hill, G.~J., Gebhardt, K., et al.\ 2011, \apj, 729, 140 


\bibitem[Finkelstein et al.(2012)]{Finkelstein2012a} Finkelstein, S.~L., 
Papovich, C., Ryan, R.~E., et al.\ 2012, \apj, 758, 93 


\bibitem[Finkelstein et al.(2012)]{Finkelstein2012b} Finkelstein, S.~L., 
Papovich, C., Salmon, B., et al.\ 2012, \apj, 756, 164 


\bibitem[Finkelstein et al.(2015)]{Finkelstein2015a} Finkelstein, S.~L., 
Ryan, R.~E., Jr., Papovich, C., et al.\ 2015, \apj, 810, 71 


\bibitem[Fontana et al.(2010)]{Fontana2010a} Fontana, A., Vanzella, E., 
Pentericci, L., et al.\ 2010, \apjl, 725, L205 


\bibitem[Furusawa et al.(2016)]{Furusawa2016a} Furusawa, H., Kashikawa, N., 
Kobayashi, M.~A.~R., et al.\ 2016, \apj, 822, 46 


\bibitem[{Geweke(1992)}]{Geweke1992a} Geweke, J. 1992, Statistical Science, 7, 94


\bibitem[Giallongo et al.(2015)]{Giallongo2015a} Giallongo, E., Grazian, A., 
Fiore, F., et al.\ 2015, \aap, 578, A83 


\bibitem[Gronwall et al.(2007)]{Gronwall2007a} Gronwall, C., Ciardullo, R., 
Hickey, T., et al.\ 2007, \apj, 667, 79 


\bibitem[Guaita et al.(2010)]{Guaita2010a} Guaita, L., Gawiser, E., Padilla, 
N., et al.\ 2010, \apj, 714, 255 


\bibitem[Hashimoto et al.(2017)]{Hashimoto2017a} Hashimoto, T., Garel, T., 
Guiderdoni, B., et al.\ 2017, \aap, 608, A10 


\bibitem[{{Hastings}(1970)}]{Hastings1970a}{Hastings}, W.~K. 1970, Biometrika, 57, 97


\bibitem[Herenz et al.(2017)]{Herenz2017a} Herenz, E.~C., Urrutia, T., 
Wisotzki, L., et al.\ 2017, \aap, 606, A12 


\bibitem[Horne(1986)]{Horne1986a} Horne, K.\ 1986, \pasp, 98, 609 


\bibitem[Hu et al.(2010)]{Hu2010a} Hu, E.~M., Cowie, L.~L., Barger, A.~J., 
et al.\ 2010, \apj, 725, 394 


\bibitem[Jung et al.(2017)]{Jung2017a} Jung, I., Finkelstein, S.~L., Song, 
M., et al.\ 2017, \apj, 834, 81 


\bibitem[Kakiichi 
\& Dijkstra(2017)]{Kakiichi2017a} Kakiichi, K., \& Dijkstra, M.\ 2017, arXiv:1710.10053 


\bibitem[Kashikawa et al.(2011)]{Kashikawa2011a} Kashikawa, N., Shimasaku, 
K., Matsuda, Y., et al.\ 2011, \apj, 734, 119 


\bibitem[Katz et al.(2018)]{Katz2018a} Katz, H., Kimm, T., Haehnelt, M., et 
al.\ 2018, arXiv:1802.01586 


\bibitem[Kimm 
\& Cen(2014)]{Kimm2014a} Kimm, T., \& Cen, R.\ 2014, \apj, 788, 121 


\bibitem[Kimm 
\& Cen(2013)]{Kimm2013a} Kimm, T., \& Cen, R.\ 2013, \apj, 776, 35 


\bibitem[Kimm et al.(2017)]{Kimm2017a} Kimm, T., Katz, H., Haehnelt, M., et 
al.\ 2017, \mnras, 466, 4826 


\bibitem[Kriek et al.(2015)]{Kriek2015a} Kriek, M., Shapley, A.~E., Reddy, 
N.~A., et al.\ 2015, \apjs, 218, 15 


\bibitem[Kurucz(1993)]{Kurucz1993a} Kurucz, R.~L.\ 1993, Kurucz CD-ROM, 
Cambridge, MA: Smithsonian Astrophysical Observatory, |c1993, December 4, 
1993,  


\bibitem[Larson et al.(2011)]{Larson2011a} Larson, D., Dunkley, J., Hinshaw, 
G., et al.\ 2011, \apjs, 192, 16 


\bibitem[Larson et al.(2017)]{Larson2017a} Larson, R.~L., Finkelstein, 
S.~L., Pirzkal, N., et al.\ 2017, arXiv:1712.05807 


\bibitem[Laursen, Sommer-Larsen, 
\& Razoumov(2011)]{Laursen2011a} Laursen, P., Sommer-Larsen, J., \& Razoumov, A.~O.\ 2011, \apj, 728, 52 


\bibitem[Livermore, Finkelstein, 
\& Lotz(2017)]{Livermore2017a} Livermore, R.~C., Finkelstein, S.~L., \& Lotz, J.~M.\ 2017, \apj, 835, 113 


\bibitem[Malhotra 
\& Rhoads(2004)]{Malhotra2004a} Malhotra, S., \& Rhoads, J.~E.\ 2004, \apjl, 617, L5 


\bibitem[Mallery et al.(2012)]{Mallery2012a} Mallery, R.~P., Mobasher, B., 
Capak, P., et al.\ 2012, \apj, 760, 128 


\bibitem[Markwardt(2009)]{Markwardt2009a} Markwardt, C.~B.\ 2009, 
Astronomical Data Analysis Software and Systems XVIII, 411, 251 


\bibitem[Marrone et al.(2018)]{Marrone2018a} Marrone, D.~P., Spilker, J.~S., 
Hayward, C.~C., et al.\ 2018, \nat, 553, 51 


\bibitem[Mason et al.(2017)]{Mason2017a} Mason, C.~A., Treu, T., Dijkstra, 
M., et al.\ 2017, arXiv:1709.05356 


\bibitem[McGreer, Mesinger, 
\& D'Odorico(2015)]{McGreer2015a} McGreer, I.~D., Mesinger, A., \& D'Odorico, V.\ 2015, \mnras, 447, 499 


\bibitem[McLure et al.(2013)]{McLure2013a} McLure, R.~J., Dunlop, J.~S., 
Bowler, R.~A.~A., et al.\ 2013, \mnras, 432, 2696 


\bibitem[{{Metropolis} {et~al.}(1953){Metropolis}, {Rosenbluth}, {Rosenbluth}, {Teller}, \& {Teller}}]{Metropolis1953a} {Metropolis}, N., {Rosenbluth}, A.~W., {Rosenbluth}, M.~N., {Teller}, A.~H., \& {Teller}, E. 1953, \jcp, 21, 1087


\bibitem[Mitra, Choudhury, 
\& Ferrara(2018)]{Mitra2018a} Mitra, S., Choudhury, T.~R., \& Ferrara, A.\ 2018, \mnras, 473, 1416 


\bibitem[Mortlock et al.(2011)]{Mortlock2011a} Mortlock, D.~J., Warren, 
S.~J., Venemans, B.~P., et al.\ 2011, \nat, 474, 616 


\bibitem[Nakajima et al.(2013)]{Nakajima2013a} Nakajima, K., Ouchi, M., 
Shimasaku, K., et al.\ 2013, \apj, 769, 3 


\bibitem[Newman et al.(2013)]{Newman2013a} Newman, J.~A., Cooper, M.~C., 
Davis, M., et al.\ 2013, \apjs, 208, 5 


\bibitem[Nilsson et al.(2009)]{Nilsson2009a} Nilsson, K.~K., Tapken, C., 
M{\o}ller, P., et al.\ 2009, \aap, 498, 13 


\bibitem[Oke 
\& Gunn(1983)]{Oke1983a} Oke, J.~B., \& Gunn, J.~E.\ 1983, \apj, 266, 713 


\bibitem[Ono et al.(2012)]{Ono2012a} Ono, Y., Ouchi, M., Mobasher, B., et 
al.\ 2012, \apj, 744, 83 


\bibitem[Ota et al.(2017)]{Ota2017a} Ota, K., Iye, M., Kashikawa, N., et 
al.\ 2017, \apj, 844, 85 


\bibitem[Ota et al.(2008)]{Ota2008a} Ota, K., Iye, M., Kashikawa, N., et 
al.\ 2008, \apj, 677, 12-26 


\bibitem[Ouchi et al.(2018)]{Ouchi2018a} Ouchi, M., Harikane, Y., Shibuya, 
T., et al.\ 2018, \pasj, 70, S13 


\bibitem[Ouchi et al.(2008)]{Ouchi2008a} Ouchi, M., Shimasaku, K., Akiyama, 
M., et al.\ 2008, \apjs, 176, 301-330 


\bibitem[Ouchi et al.(2010)]{Ouchi2010a} Ouchi, M., Shimasaku, K., Furusawa, 
H., et al.\ 2010, \apj, 723, 869 


\bibitem[Paardekooper, Khochfar, 
\& Dalla Vecchia(2015)]{Paardekooper2015a} Paardekooper, J.-P., Khochfar, S., \& Dalla Vecchia, C.\ 2015, \mnras, 451, 2544 


\bibitem[Papovich et al.(2011)]{Papovich2011a} Papovich, C., Finkelstein, 
S.~L., Ferguson, H.~C., Lotz, J.~M., 
\& Giavalisco, M.\ 2011, \mnras, 412, 1123 


\bibitem[Pentericci et al.(2011)]{Pentericci2011a} Pentericci, L., Fontana, 
A., Vanzella, E., et al.\ 2011, \apj, 743, 132 


\bibitem[Pentericci et al.(2014)]{Pentericci2014a} Pentericci, L., Vanzella, 
E., Fontana, A., et al.\ 2014, \apj, 793, 113 


\bibitem[Pirzkal et al.(2004)]{Pirzkal2004a} Pirzkal, N., Xu, C., Malhotra, 
S., et al.\ 2004, \apjs, 154, 501 


\bibitem[Pirzkal et al.(2017)]{Pirzkal2017a} Pirzkal, N., Malhotra, S., 
Ryan, R.~E., et al.\ 2017, \apj, 846, 84 


\bibitem[Planck Collaboration et al.(2016)]{Planck-Collaboration2016a} 
Planck Collaboration, Ade, P.~A.~R., Aghanim, N., et al.\ 2016, \aap, 594, 
A13 


\bibitem[Pradhan, Montenegro, Nahar, 
\& Eissner(2006)]{Pradhan2006a} Pradhan, A.~K., Montenegro, M., Nahar, S.~N., \& Eissner, W.\ 2006, \mnras, 366, L6 


\bibitem[Rhoads et al.(2005)]{Rhoads2005a} Rhoads, J.~E., Panagia, N., 
Windhorst, R.~A., et al.\ 2005, \apj, 621, 582 


\bibitem[{Roberts {et~al.}(1997)Roberts, Gelman, \& Gilks}]{Roberts1997a} Roberts, G.~O., Gelman, A., \& Gilks, W.~R. 1997, The Annals of Applied Probability, 7, 110


\bibitem[Robertson, Ellis, Furlanetto, 
\& Dunlop(2015)]{Robertson2015a} Robertson, B.~E., Ellis, R.~S., Furlanetto, S.~R., \& Dunlop, J.~S.\ 2015, \apjl, 802, L19 


\bibitem[Robertson et al.(2013)]{Robertson2013a} Robertson, B.~E., 
Furlanetto, S.~R., Schneider, E., et al.\ 2013, \apj, 768, 71 


\bibitem[Rosdahl et al.(2018)]{Rosdahl2018a} Rosdahl, J., Katz, H., Blaizot, 
J., et al.\ 2018, arXiv:1801.07259 


\bibitem[Salmon et al.(2015)]{Salmon2015a} Salmon, B., Papovich, C., 
Finkelstein, S.~L., et al.\ 2015, \apj, 799, 183 


\bibitem[Salpeter(1955)]{Salpeter1955a} Salpeter, E.~E.\ 1955, \apj, 121, 
161 


\bibitem[Schaerer, de Barros, 
\& Stark(2011)]{Schaerer2011a} Schaerer, D., de Barros, S., \& Stark, D.~P.\ 2011, \aap, 536, A72 


\bibitem[Schenker, Ellis, Konidaris, 
\& Stark(2014)]{Schenker2014a} Schenker, M.~A., Ellis, R.~S., Konidaris, N.~P., \& Stark, D.~P.\ 2014, \apj, 795, 20 


\bibitem[Schenker et al.(2013)]{Schenker2013a} Schenker, M.~A., Robertson, 
B.~E., Ellis, R.~S., et al.\ 2013, \apj, 768, 196 


\bibitem[Schenker et al.(2012)]{Schenker2012a} Schenker, M.~A., Stark, 
D.~P., Ellis, R.~S., et al.\ 2012, \apj, 744, 179 


\bibitem[Schmidt et al.(2016)]{Schmidt2016a} Schmidt, K.~B., Treu, T., 
Brada{\v c}, M., et al.\ 2016, \apj, 818, 38 


\bibitem[Song et al.(2016)]{Song2016a} Song, M., Finkelstein, S.~L., Ashby, 
M.~L.~N., et al.\ 2016, \apj, 825, 5 


\bibitem[Song et al.(2014)]{Song2014a} Song, M., Finkelstein, S.~L., 
Gebhardt, K., et al.\ 2014, \apj, 791, 3 


\bibitem[Song et al.(2016)]{Song2016b} Song, M., Finkelstein, S.~L., 
Livermore, R.~C., et al.\ 2016, \apj, 826, 113 


\bibitem[Stark et al.(2009)]{Stark2009a} Stark, D.~P., Ellis, R.~S., Bunker, 
A., et al.\ 2009, \apj, 697, 1493 


\bibitem[Stark et al.(2010)]{Stark2010a} Stark, D.~P., Ellis, R.~S., Chiu, 
K., Ouchi, M., \& Bunker, A.\ 2010, \mnras, 408, 1628 


\bibitem[Stark, Ellis, 
\& Ouchi(2011)]{Stark2011a} Stark, D.~P., Ellis, R.~S., \& Ouchi, M.\ 2011, \apjl, 728, L2 


\bibitem[Szokoly(2005)]{Szokoly2005a} Szokoly, G.~P.\ 2005, \aap, 443, 703 


\bibitem[Tilvi et al.(2014)]{Tilvi2014a} Tilvi, V., Papovich, C., 
Finkelstein, S.~L., et al.\ 2014, \apj, 794, 5 


\bibitem[Trainor, Strom, Steidel, 
\& Rudie(2016)]{Trainor2016a} Trainor, R.~F., Strom, A.~L., Steidel, C.~C., \& Rudie, G.~C.\ 2016, \apj, 832, 171 


\bibitem[Treu et al.(2013)]{Treu2013a} Treu, T., Schmidt, K.~B., Trenti, M., 
Bradley, L.~D., \& Stiavelli, M.\ 2013, \apjl, 775, L29 


\bibitem[Treu et al.(2012)]{Treu2012a} Treu, T., Trenti, M., Stiavelli, M., 
Auger, M.~W., \& Bradley, L.~D.\ 2012, \apj, 747, 27 


\bibitem[van Dokkum(2001)]{van-Dokkum2001a} van Dokkum, P.~G.\ 2001, \pasp, 
113, 1420 


\bibitem[Vanzella et al.(2014)]{Vanzella2014a} Vanzella, E., Fontana, A., 
Pentericci, L., et al.\ 2014, \aap, 569, A78 


\bibitem[Wold, Barger, 
\& Cowie(2014)]{Wold2014a} Wold, I.~G.~B., Barger, A.~J., \& Cowie, L.~L.\ 2014, \apj, 783, 119 


\bibitem[Wold et al.(2017)]{Wold2017a} Wold, I.~G.~B., Finkelstein, S.~L., 
Barger, A.~J., Cowie, L.~L., \& Rosenwasser, B.\ 2017, \apj, 848, 108 


\bibitem[Worseck et al.(2014)]{Worseck2014a} Worseck, G., Prochaska, J.~X., 
O'Meara, J.~M., et al.\ 2014, \mnras, 445, 1745 


\bibitem[Yoshiura et al.(2017)]{Yoshiura2017a} Yoshiura, S., Hasegawa, K., 
Ichiki, K., et al.\ 2017, \mnras, 471, 3713 


\bibitem[Zheng et al.(2014)]{Zheng2014a} Zheng, Z.-Y., Wang, J.-X., 
Malhotra, S., et al.\ 2014, \mnras, 439, 1101 


\bibitem[Zheng et al.(2017)]{Zheng2017a} Zheng, Z.-Y., Wang, J., Rhoads, J., 
et al.\ 2017, \apjl, 842, L22 

\end{thebibliography}
\end{document}